\documentclass[english]{article}
\usepackage[T1]{fontenc}
\usepackage[latin9]{inputenc}
\usepackage{verbatim}
\usepackage{amsmath}
\usepackage{amssymb}
\usepackage{graphicx}
\usepackage{subfig}
\usepackage{multicol}

\makeatletter
\@ifundefined{date}{}{\date{}}

\makeatother

\usepackage{babel}
\begin{document}

\title{Join Query Optimization Techniques for Complex Event Processing Applications}
\maketitle \begin{multicols}{2}
\begin{center} Ilya Kolchinsky\\ Technion, Israel Institute of Technology \par\end{center}
\begin{center} Assaf Schuster\\ Technion, Israel Institute of Technology \par\end{center}
\end{multicols}
\begin{abstract}
Complex event processing (CEP) is a prominent technology used in many
modern applications for monitoring and tracking events of interest
in massive data streams. CEP engines inspect real-time information
flows and attempt to detect combinations of occurrences matching predefined
patterns. This is done by combining basic data items, also called
``primitive events'', according to a pattern detection plan, in
a manner similar to the execution of multi-join queries in traditional
data management systems. Despite this similarity, little work has
been done on utilizing existing join optimization methods to improve
the performance of CEP-based systems.

In this paper, we provide the first theoretical and experimental study
of the relationship between these two research areas. We formally
prove that the CEP Plan Generation problem is equivalent to the Join
Query Plan Generation problem for a restricted class of patterns and
can be reduced to it for a considerably wider range of classes. This
result implies the NP-completeness of the CEP Plan Generation problem.
We further show how join query optimization techniques developed over
the last decades can be adapted and utilized to provide practically
efficient solutions for complex event detection. Our experiments demonstrate
the superiority of these techniques over existing strategies for CEP
optimization in terms of throughput, latency, and memory consumption.
\end{abstract}

\section{Introduction}

\label{sec:Introduction}

Complex event processing has become increasingly important for applications
in which arbitrarily complex patterns must be efficiently detected
over high-speed streams of events. Online finance, security monitoring,
and fraud detection are among the many examples. Pattern detection
generally consists of collecting primitive events and combining them
into potential (partial) matches using some type of detection model.
As more events are added to a partial match, a full pattern match
is eventually formed and reported. Popular CEP mechanisms include
nondeterministic finite automata (NFAs) \cite{AgrawalDGI2008,DemersJB07,WuDR06},
finite state machines \cite{AkdereMCT08,Schultz-MollerMP09}, trees
\cite{MeiM09}, and event processing networks \cite{EtzionN10,RabinovichEG11}.

A CEP engine creates an internal representation for each pattern $P$
to be monitored. This representation is based on a model used for
detection (e.g., an automaton or a tree) and reflects the structure
of $P$. In some systems \cite{AgrawalDGI2008,WuDR06}, the translation
from a pattern specification to a corresponding representation is
a one-to-one mapping. Other frameworks \cite{AkdereMCT08,KolchinskySS15,MeiM09,RabinovichEG11,Schultz-MollerMP09}
introduce the notion of a \textit{cost-based evaluation plan}, where
multiple representations of $P$ are possible, and one is chosen according
to the user's preference or some predefined cost metric.

We will illustrate the above using the following example. Assume that
we are receiving periodical readings from four traffic cameras \textit{A},
\textit{B}, \textit{C} and \textit{D}. We are required to recognize
a sequence of appearances of a particular vehicle on all four cameras
in order of their position on a road, e.g., \textit{$A\rightarrow B\rightarrow C\rightarrow D$}.
Assume also that, due to a malfunction in camera \textit{D}, it only
transmits one frame for each 10 frames sent by the other cameras.

\begin{figure}
	\centering
	\subfloat[]{\includegraphics[width=.6\linewidth]{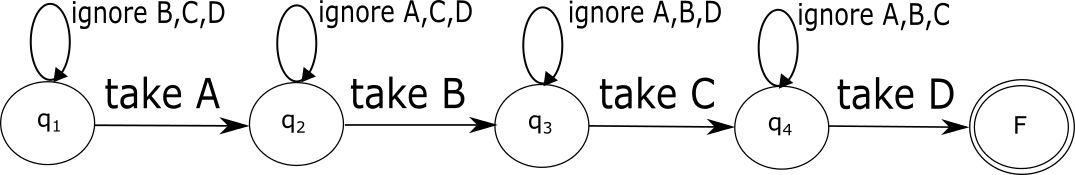}\label{fig:nfa-no-reordering}}\\[2ex]
	\subfloat[]{\includegraphics[width=.6\linewidth]{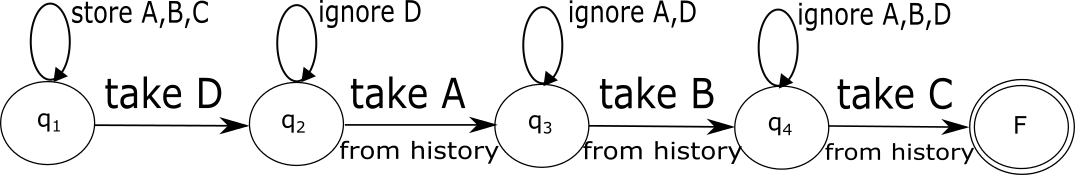}\label{fig:nfa-with-reordering}}\\[2ex] 	\subfloat[]{\includegraphics[width=.3\linewidth]{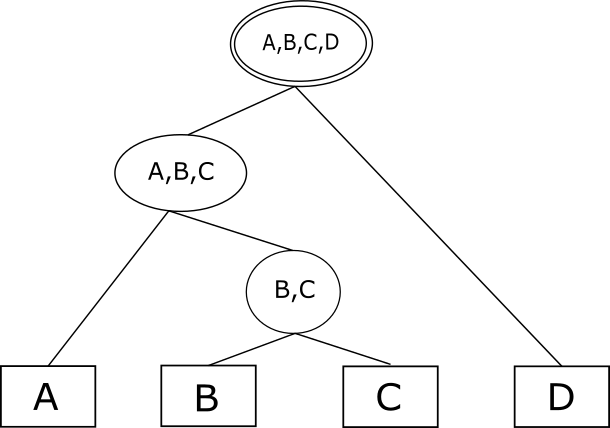}\label{fig:zstream-abcd}}
    \caption{Evaluation structures for a CEP pattern \textit{SEQ(A,B,C,D)}: \protect\subref{fig:nfa-no-reordering} NFA with no reordering; \protect\subref{fig:nfa-with-reordering} NFA with reordering; \protect\subref{fig:zstream-abcd} evaluation tree.}
	\label{fig:cep-evaluation}
\end{figure}

Figure \ref{fig:cep-evaluation}\subref{fig:nfa-no-reordering} displays
a nondeterministic finite automaton (NFA) for detecting this pattern,
as described in \cite{WuDR06}. A state is defined for each prefix
of a valid match. During evaluation, a combination of camera readings
matching each prefix will be represented by a unique instance of the
NFA in the corresponding state. Transitions between states are triggered
nondeterministically by the arrival of an event satisfying the constraints
defined by the pattern. A new NFA instance is created upon each transition.

The structure of the above automaton is uniquely dictated by the order
of events in the given sequence. However, due to the low transmission
rate of \textit{D}, it would be beneficial to wait for its signal
before examining the local history for previous readings of \textit{A},
\textit{B} and \textit{C} that match the constraints. This way, fewer
prefixes would be created. Figure \ref{fig:cep-evaluation}\subref{fig:nfa-with-reordering}
demonstrates an out-of-order NFA for the rewritten pattern (defined
as ``Lazy NFA'' in \cite{KolchinskySS15}). It starts by monitoring
the rarest event type \textit{D} and storing the other events in the
dedicated buffer. As a reading from camera \textit{D} arrives, the
buffer is inspected for events from \textit{A}, \textit{B} and \textit{C}
preceding the one received from \textit{D} and located in the same
time window. This plan is more efficient than the one implicitly used
by the first automaton in terms of the total number of partial matches
created during evaluation. Moreover, unless more constraints on the
events are defined, it is the cheapest among all $\left(4!\right)$
available plans, that is, all mutual orders of \textit{A}, \textit{B},
\textit{C} and \textit{D}.

Not all CEP mechanisms represent a plan as an evaluation order. In
Figure \ref{fig:cep-evaluation}\subref{fig:zstream-abcd} a tree-based
evaluation mechanism \cite{MeiM09} for detecting the above pattern
is depicted. Events are accepted at the corresponding leaves of the
tree, and passed towards the root where full matches are reported.
Note that this model requires an evaluation plan to be supplied, because,
for a pattern of size $n$, there are at least $C_{n-1}=\frac{\left(2n-2\right)!}{\left(n-1\right)!n!}$
possible trees (where $C_{n}$ is the $n^{th}$ Catalan number).

In many scenarios, we will prefer the evaluation mechanisms supporting
cost-based plan generation over those mechanisms allowing for only
one such plan to be defined. This way, we can drastically boost system
performance subject to selected metrics by picking more efficient
plans. However, as the space of potential plans is at least exponential
in pattern size, finding an optimal evaluation plan is not a trivial
task.

Numerous authors have identified and targeted this issue. Some of
the proposed solutions are based on rewriting the original pattern
according to a set of predefined rules to maximize the efficiency
of its detection \cite{RabinovichEG11,Schultz-MollerMP09}. Other
approaches discuss various strategies and algorithms for generating
an evaluation plan that maximizes the performance for a given pattern
according to some cost function \cite{AkdereMCT08,KolchinskySS15,MeiM09}.
While the above approaches demonstrate promising results, this research
field remains largely unexplored, and the space of the potential optimization
techniques is still far from being exhausted.

\begin{figure}
	\centering
	\subfloat[]{\includegraphics[width=.45\linewidth]{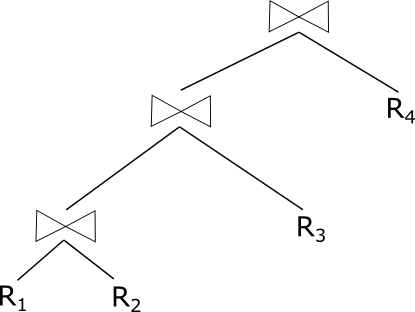}\label{fig:left-deep-tree}}
	\quad\quad
	\subfloat[]{\includegraphics[width=.45\linewidth]{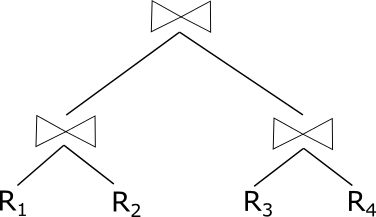}\label{fig:bushy-tree}}
    \caption{Execution plans for a join of four relations \textit{$R_{1},R_{2},R_{3},R_{4}$}: \protect\subref{fig:left-deep-tree} a left-deep tree; \protect\subref{fig:bushy-tree} a bushy tree.}
	\label{fig:join-trees}
\end{figure}

The problem described above closely resembles the problem of estimating
execution plans for large join queries. As opposed to CEP plan generation,
this is a well-known, established, and extensively targeted research
topic. A plethora of methods and approaches producing close-to-optimal
results were published during the last few decades. These methods
range from simple greedy heuristics, to exhaustive dynamic programming
techniques, to randomized and genetic algorithms \cite{KrishnamurthyBZ86,LeeSC97,MoerkotteN06,SelingerACLP79,SteinbrunnMK97,Swami89}.
Figure \ref{fig:join-trees} illustrates two main types of execution
plans for join queries, a left-deep tree (\ref{fig:join-trees}\subref{fig:left-deep-tree})
and a bushy tree (\ref{fig:join-trees}\subref{fig:bushy-tree})  \cite{IoannidisK91}.

Both problems look for a way to efficiently combine multiple data
items such that some cost function is minimized. Also, both produce
solutions possessing similar structures. If we reexamine Figures \ref{fig:cep-evaluation}
and \ref{fig:join-trees}, we can see that left-deep tree plans (\ref{fig:join-trees}\subref{fig:left-deep-tree})
and bushy tree plans (\ref{fig:join-trees}\subref{fig:bushy-tree})
closely resemble evaluation plans for NFAs (\ref{fig:cep-evaluation}\subref{fig:nfa-with-reordering})
and trees (\ref{fig:cep-evaluation}\subref{fig:zstream-abcd}) respectively.
An interesting question is whether join-related techniques can be
used to create better CEP plans using a proper reduction.

In this work, we attempt to close the gap between the two areas of
research. We study the relationship between CEP Plan Generation (CPG)
and Join Query Plan Generation (JQPG) problems and show that any instance
of CPG can be transformed into an instance of JQPG. Consequently,
any existing method for JQPG can be made applicable to CPG. Our contributions
can be summarized as follows:

\textbullet{} We formally prove the equivalence of JQPG and CPG for
a large subset of CEP patterns, the conjunctive patterns. The proof
addresses the two major classes of evaluation plans, the order-based
plans and the tree-based plans (Section \ref{sec:The-Equivalence-of}).

\textbullet{} We extend the above result by showing how other pattern
types can be converted to conjunctive patterns, thus proving that
any instance of CPG can be reduced to an instance of JQPG (Section
\ref{sec:General-Pattern-Types}).

\textbullet{} The deployment of a JQPG method to CPG is not trivial,
as multiple CEP-specific issues need to be addressed, such as detection
latency constraints, event consumption policies, and adaptivity considerations.
We present and discuss the steps essential for successful adaptation
of JQPG techniques to the CEP domain (Section \ref{sec:Adapting-Join-Query}).

\textbullet{} We validate our theoretical analysis in an extensive
experimental study. Several well-known JQPG methods, such as Iterative
Improvement \cite{Swami89} and Dynamic Programming \cite{SelingerACLP79},
were applied on a real-world event dataset and compared to the existing
state-of-the-art CPG mechanisms. The results demonstrate the superiority
of the adapted JQPG techniques (Section \ref{sec:Experimental-Evaluation}).

\section{Complex Event Processing}

\label{sec:Background-and-Terminology}

In this section, we introduce the notations used throughout this paper
and provide the necessary background on complex event processing.
We present the elements of a CEP pattern, including a brief taxonomy
of commonly used pattern types. Then, we describe two classes of CEP
evaluation mechanisms that are the focus of this work: the order-based
and the tree-based CEP. The results obtained in the later sections
are based on but not limited to these two representation models, and
they can be extended to more complex schemes, such as event processing
networks \cite{EtzionN10,RabinovichEG11}.

\subsection{CEP Patterns}

\label{sub:CEP-Pattern-Types}

The patterns recognized by CEP systems are normally formed using declarative
specification languages \cite{CugolaM10,DemersJB07,WuDR06}. A pattern
is defined by a combination of primitive events, operators, a set
of predicates to be satisfied by the participating events, and a time
window. Each event is represented by a type and a set of attributes,
including the occurrence timestamp. The operators describe the relations
between different events comprising a pattern match. The predicates,
usually organized in a Boolean formula, specify the constraints on
the attribute values of the events. As an example, consider the following
pattern specification syntax, taken from SASE \cite{WuDR06}:

\[
\begin{array}{l}
PATTERN\: op\left(T_{1}\: e_{1},T_{2}\: e_{2},\cdots,T_{n}\: e_{n}\right)\\
WHERE\:(c_{1,1}\wedge c_{1,2}\wedge\cdots\wedge c_{n,n-1}\wedge c_{n,n})\\
WITHIN\: W.
\end{array}
\]%

Here, the PATTERN clause specifies the events $e_{1},\cdots,e_{n}$
we would like to detect and the operator $op$ to combine them (see
below). The WHERE clause defines a Boolean CNF formula of inter-event
constraints, where $c_{i,j};1\leq i,j\leq n$ stands for the mutual
condition between attributes of $e_{i}$ and $e_{j}$. $c_{i,i}$
declares filter conditions on $e_{i}$. Any of $c_{i,j}$ can be empty.
For the rest of our paper, we assume that all conditions between events
are at most pairwise (i.e., a single condition involves at most two
different events). This assumption is for presentational purposes
only, as our results can be easily generalized to arbitrary predicates.
The WITHIN clause sets the time window $W$, which is the maximal
allowed difference between the timestamps of any pair of events in
a match.

Throughout this work we assume that each primitive event has a well-defined
type, i.e., the event either contains the type as an attribute or
it can be easily inferred from other attributes using negligible system
resources. While this constraint may seem limiting, it is easy to
overcome in most cases by redefining what a pattern creator considers
a type.

In this paper, we will consider the most commonly used operators,
namely AND, SEQ, and OR. The AND operator requires the occurrence
of all events specified in the pattern within the time window. The
SEQ operator extends this definition by also expecting the events
to appear in a predefined temporal order. The OR operator corresponds
to the appearance of any event out of those specified.

Two additional operators of particular importance are the negation
operator (NOT) and the Kleene closure operator (KL). They can only
be applied on a single event and are used in combination with other
operators. $NOT\left(e_{i}\right)$ requires the absence of the event
$e_{i}$ from the stream (or from a specific position in the pattern
in the case of the SEQ operator), whereas $KL\left(e_{i}\right)$
accepts one or more instances of $e_{i}$. In the remainder of this
paper, we will refer to NOT and KL as \textit{unary operators}, while
AND, SEQ and OR will be called \textit{n-ary operators}.

The PATTERN clause may include an unlimited number of n-ary and unary
operators. We will refer to patterns containing a single n-ary operator,
and at most a single unary operator per primitive event, as \textit{simple
patterns}. On the contrary, \textit{nested patterns} are allowed to
contain multiple n-ary operators (e.g., a disjunction of conjunctions
and/or sequences will be considered a nested pattern). Nested patterns
present an additional level of complexity and require advanced techniques
(e.g., as described in \cite{LiuRDGWAM11}).

We will further divide simple patterns into subclasses. A simple pattern
whose n-ary operator is an AND operator will be denoted as a \textit{conjunctive
pattern}. Similarly, \textit{sequence pattern} and \textit{disjunctive
pattern} will stand for patterns with SEQ and OR operators, respectively.
In addition, a simple pattern containing no unary operators will be
called a \textit{pure pattern}. 

The ``four cameras pattern'' described in Section \ref{sec:Introduction}
illustrates the above. This is a pure sequence pattern, written in
SASE as follows:

\[
\begin{array}{l}
PATTERN\: SEQ\left(A\: a,B\: b,C\: c,D\: d\right)\\
WHERE\:(a.vehicleID=b.vehicleID=\\
\qquad \qquad =c.vehicleID=d.vehicleID)\\
WITHIN\: W.
\end{array}
\]%

The following example depicts a nested pattern, consisting of a (non-pure)
conjunctive pattern and an inner pure disjunctive pattern:

\[
\begin{array}{l}
PATTERN\: AND\left(A\: a,NOT\left(B\: b\right),OR\left(C\: c,D\: d\right)\right)\\
WITHIN\: W.
\end{array}
\]%

\subsection{Order-based Evaluation Mechanisms}

\label{sub:Order-based-Evaluation-Mechanism}

Order-based evaluation mechanisms play an important role in CEP engines
based on state machines. One of the most commonly used models following
this principle is the NFA (nondeterministic finite automaton) \cite{AgrawalDGI2008,DemersJB07,WuDR06}.
An NFA consists of a set of states and conditional transitions between
them. Each state corresponds to a prefix of a full pattern match.
Transitions are triggered by the arrival of the primitive events,
which are then added to partial matches. Figures \ref{fig:cep-evaluation}\subref{fig:nfa-no-reordering}
and \ref{fig:cep-evaluation}\subref{fig:nfa-with-reordering} depict
an example of two NFAs constructed for the same sequence pattern using
different order-based plans. While in theory NFAs may possess an arbitrary
topology, non-nested patterns are normally detected by a chain-like
structure.

The basic NFA model does not include any notion of altering the ``natural''
evaluation order or any other optimization based on pattern rewriting.
Multiple works have presented methods for constructing NFAs with out-of-order
processing support. W.l.o.g., we will use the Lazy NFA mechanism,
a chain-structured NFA introduced in \cite{KolchinskySK17Long,KolchinskySS15}
and capable of following a specified evaluation order.

Given a pattern of $n$ events and a user-specified order $O$ on
the event types appearing in the pattern, a chain of $n+1$ states
is constructed, with each state $k$ corresponding to a match prefix
of size $k-1$. The order of the states matches $O$ (i.e., the transition
from the initial state to the second one expects the first type in
$O$, the transition from the second to the third state expects the
second type in $O$, etc.). The $\left(n+1\right)^{th}$ state in
the chain is the accepting state. To achieve out-of-order evaluation,
NFA instances store events which arrive out-of-order. A buffered event
is retrieved and processed when its corresponding state in the chain
is reached. During a traversal attempt on an edge connecting two adjacent
states $q_{i}$ and $q_{i+1}$, the following conditions will be verified:
$\left\{ c_{i+1,j}|e_{j}\: precedes\: e_{i+1}\: in\: O\right\} $.

This construction method allows us to apply all possible $\left(n!\right)$
orders. Note that the detection correctness is not affected, i.e.,
all $\left(n!\right)$ NFAs will track the exact same pattern.

\subsection{Tree-based Evaluation Mechanisms}

\label{sub:Tree-based-Evaluation-Mechanisms}

An alternative to NFA, the tree-based evaluation mechanism \cite{MeiM09}
specifies which subsets of full pattern matches are to be tracked
by defining tree-like structures. For each event participating in
a pattern, a designated leaf is created. During evaluation, events
are routed to their corresponding leaves and are buffered there. The
non-leaf nodes accumulate the partial matches. The computation at
each non-leaf node proceeds only when all of its children are available
(i.e., all events have arrived or partial matches have been calculated).
Matches formed at the tree root are reported to the end users. An
example is shown in Figure \ref{fig:cep-evaluation}\subref{fig:zstream-abcd}.

The main advantage of the tree-based evaluation mechanism over NFA
is its flexibility. Instead of relying on a single evaluation order,
it allows primitive events to arrive in any possible order and still
be efficiently processed.

ZStream assumed a batch-iterator setting \cite{MeiM09}. To perform our 
study under a unified framework, we modify this behavior to support 
arbitrary time windows. As described above with regard to NFAs, a separate tree instance
will be created for each currently found partial match. As a new event
arrives, an instance will be created containing this event. Every
instance $I$ corresponds to some subtree $s$ of the tree plan, with
the leaves of $s$ holding the primitive events in $I$. Whenever
a new instance $I'$ is created, the system will attempt to combine
it with previously created ``siblings'', that is, instances corresponding
to the subtree sharing the parent node with $s'$. As a result, another
new instance containing the unified subtree will be generated. This
in turn will trigger the same process again, and it will proceed recursively
until the root of the tree is reached or no siblings are found.

Contrary to the basic version of NFA \cite{AgrawalDGI2008,DemersJB07,WuDR06}, 
ZStream includes an algorithm for determining the
optimal tree structure for a given pattern. This algorithm is based
on a cost model that takes into account the arrival rates of the primitive
events and the selectivities of their predicates. However, since leaf
reordering is not supported, a subset of potential plans is missed.
We will illustrate this drawback using the following example:

\[
\begin{array}{l}
PATTERN\: SEQ\left(A\: a,B\: b,C\: c\right)\\
WHERE\:(a.x=c.x)\\
WITHIN\: W.
\end{array}
\]%

\begin{figure*}
	\centering
	\subfloat[]{\includegraphics[width=.2\linewidth]{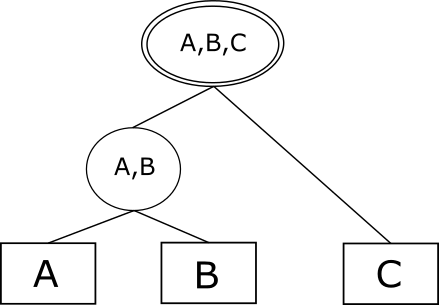}\label{fig:zstream-abc-left}}\quad\quad\quad 	\subfloat[]{\includegraphics[width=.2\linewidth]{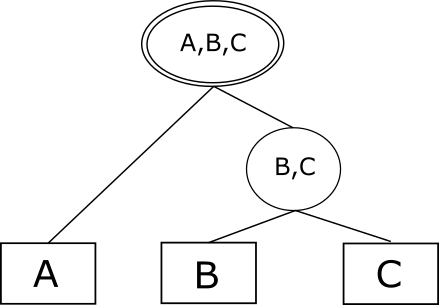}\label{fig:zstream-abc-right}}\quad\quad\quad 	\subfloat[]{\includegraphics[width=.2\linewidth]{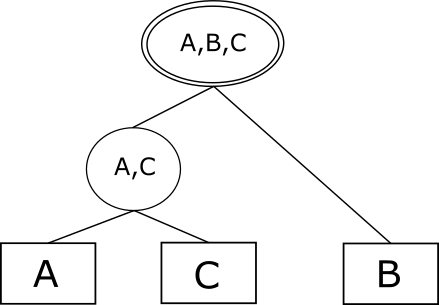}\label{fig:zstream-abc-reordering}}
    \caption{Evaluation trees for a pattern \textit{SEQ(A,B,C)}: \protect\subref{fig:zstream-abc-left} a left-deep tree produced by ZStream; \protect\subref{fig:zstream-abc-right} a right-deep tree produced by ZStream; \protect\subref{fig:zstream-abc-reordering} an optimal evaluation tree, which cannot be produced by ZStream.}
	\label{fig:zstream-reordering-example}
\end{figure*}

We assume that all events arrive at identical rates, and that the
condition between \textit{A} and \textit{C} is very restrictive. Figures
\ref{fig:zstream-reordering-example}\subref{fig:zstream-abc-left}
and \ref{fig:zstream-reordering-example}\subref{fig:zstream-abc-right}
present the only two possible plans according to the algorithm presented
in \cite{MeiM09}. However, due to the condition between \textit{A}
and \textit{C}, the most efficient evaluation plan is the one displayed
in Figure \ref{fig:zstream-reordering-example}\subref{fig:zstream-abc-reordering}.
It will be shown later how join query optimization methods can be
incorporated to overcome this issue.

\section{Plan Generation Problems}

\label{sec:Plan-Generation-Problems}

This section defines and describes the two problems whose relationship
will be closely studied in the subsequent sections. We start by presenting
the CEP Plan Generation problem and explain its two variations, order-based
CPG and tree-based CPG. Then, we briefly outline the Join Query Plan
Generation (JQPG) problem.

\subsection{CEP Plan Generation}

\label{sub:CEP-Plan-Generation}

We will start with the definition of the CEP evaluation plan. The
\textit{evaluation plan} provides a scheme for the evaluation mechanism,
according to which its internal pattern representation is created.
Therefore, different evaluation plans are required for different CEP
frameworks. In this paper, we distinguish between two main types of
plans, the \textit{order-based plan} and the \textit{tree-based plan},
with more complex types left for future work.

An order-based plan consists of a permutation of the primitive event
types declared by the pattern. An order-based CEP engine uses this
plan to set the order in which events are processed at runtime. Order-based
plans are applicable to mechanisms evaluating a pattern event-by-event,
as described in Section \ref{sub:Order-based-Evaluation-Mechanism}.

A tree-based plan extends the above by providing a tree-like scheme
for pattern evaluation. In this scheme, the structure of the internal
nodes serves as a loose order in which different events are to be
matched. It specifies which subsets of valid matches are to be locally
buffered and how to combine them into larger partial matches. Plans
of this type can be used for the tree-based evaluation mechanism presented
in Section \ref{sub:Tree-based-Evaluation-Mechanisms}.

We can thus define two variations of the CEP Plan Generation problem,
\textit{order-based CPG} and \textit{tree-based CPG}. In each variation,
the goal is to determine an optimal evaluation plan $P$ subject to
some cost function $Cost\left(P\right)$. Different CEP systems define
different metrics to measure their efficiency. In this paper we will
consider a highly relevant performance optimization goal: reducing
the number of active partial matches within the time window (denoted
below simply as \textit{number of partial matches}).

Regardless of the system-specific performance objectives, the implicit
requirement to monitor all valid subsets of primitive events can become
a major bottleneck. Because any partial match might form a full pattern
match, their number is worst-case exponential in the number of events
participating in a pattern. Further, as a newly arrived event needs
to be checked against all (or most of) the currently stored partial
matches, the processing time and resource consumption per event can
become impractical for real-time applications. Other metrics, such
as detection latency or network communication cost, may also be negatively
affected. Thus, given the crucial role of the number of partial matches
in all aspects of CEP, it was chosen as our primary cost function.

The formal definitions of the cost functions for order-based CPG and
tree-based CPG will be given in Sections \ref{sub:Order-Based-Evaluation}
and \ref{sub:Tree-Based-Evaluation} respectively.

\subsection{Join Query Plan Generation}

\label{sub:Join-Query-Plan}

Join Query Plan Generation is a well-known problem in query optimization
\cite{KrishnamurthyBZ86,SelingerACLP79,Swami89}. In this problem,
we are given relations $R_{1},\cdots,R_{n}$ and a query graph describing
the mutual conditions to be satisfied by the tuples of the relations in
order to be included in the result. A condition between any pair of
relations $R_{i},R_{j}$ has a known selectivity $f_{i,j}\in\left[0,1\right]$
(we define $f_{i,j}=1$ if no such condition is defined between the
two). The goal is to produce a query plan for this join operation,
such that a predefined cost function defined on the plan space will
be minimized.

One popular choice for the cost function is the number of intermediate
tuples produced during plan execution. For the rest of this paper,
we will refer to it as the \textit{intermediate results size}. In
\cite{CluetM95}, the following expression is given to calculate this
function for each two-way join of two input relations:
\[
C\left(R_{i},R_{j}\right)=\left|R_{i}\right|\cdot\left|R_{j}\right|\cdot f_{i,j},
\]
where $\left|R_{i}\right|,\left|R_{j}\right|$ are the cardinalities
of the joined relations. This formula is naturally extended to relations
produced during join calculation:
\[
C\left(S,T\right)=\left|S\right|\cdot\left|T\right|\cdot f_{S,T}.
\]

Here, $S=R_{i_{1}}\bowtie\cdots\bowtie R_{i_{s}};T=R_{j_{1}}\bowtie\cdots\bowtie R_{j_{t}}$
are the partial join results of some subsets of $R_{1},\cdots,R_{n}$
and 
\[
f_{S,T}=\left|S\bowtie T\right|/\left|S\times T\right|
\]
is the product of selectivities of all predicates defined
between the individual relations comprising $S$ and $T$.

The two most popular classes of join query plans are the left-deep
trees and the bushy trees. Algorithms based on the former type restrict
their output to plans with a so-called ``left-deep'' topology. This
type of join tree processes the input relations one-by-one, adding
a new relation to the current intermediate result during each step.
Hence, for this class of techniques, a valid solution is a join order
rather than a join plan, since for any order over $R_{1},\cdots,R_{n}$
there exists exactly one left-deep tree. Figure \ref{fig:join-trees}\subref{fig:left-deep-tree}
depicts an example of a left-deep tree for a join query of four relations.

Approaches based on bushy trees pose no limitations on the plan topology,
allowing it to contain arbitrary branches. An example is shown in
Figure \ref{fig:join-trees}\subref{fig:bushy-tree}. Here a valid
solution specifies a complete join tree rather than merely an order.

In Sections \ref{sub:Order-Based-Evaluation} and \ref{sub:Tree-Based-Evaluation}
we will provide the formal definitions of the execution plan costs
for left-deep and bushy join trees, respectively.
The join query plan generation problem was shown by multiple authors
to be NP-complete \cite{CluetM95,Ibaraki84}, even when only left-deep
trees are considered.

\section{The Equivalence of CPG and JQPG for Pure Conjunctive Patterns}

\label{sec:The-Equivalence-of}

This section presents the formal proof of equivalence between CPG
and JQPG for pure conjunctive patterns. We show that, when the pattern
to be monitored is a pure conjunctive pattern and the CPG cost function
represents the number of partial matches, the two problems are equivalent.
From this result, we deduce the NP-completeness of CPG. In Section
\ref{sec:General-Pattern-Types} we demonstrate how to convert non-pure
sequence and conjunctive patterns to pure conjunctive form, thus proving
that an instance of CPG for these pattern types can be reduced to
an instance of JQPG (though the opposite does not hold).

\subsection{Order-Based Evaluation}

\label{sub:Order-Based-Evaluation}

We will first focus on a CPG variation for order-based evaluation
plans. In this section we will show that this problem is equivalent
to JQPG restricted to left-deep trees. To that end, we will define
the cost model functions for both problems and then present the equivalence
theorem.

Our cost function $Cost_{ord}$ will reflect the number of partial
matches coexisting in memory within the time window. The calculations
will be based on the arrival rates of the events and the selectivities
of the predicates.

Let $sel_{i,j}$ denote the selectivity of $c_{i,j}$, i.e., the probability
of a partial match containing instances of events of types $T_{i}$
and $T_{j}$ to pass the condition. Additionally, let $r_{1},\cdots r_{n}$
denote the arrival rates of corresponding event types $T_{1},\cdots T_{n}$.
Then, the expected number of primitive events of type $T_{i}$ arriving
within the time window $W$ is $W\cdot r_{i}$. Let $O=\left(T_{p_{1}},T_{p_{2}},\cdots T_{p_{n}}\right);p_{i}\in[1,n]$
denote an execution order. Then, during pattern evaluation according
to $O$, the expected number of partial matches of length $k,1\leq k\leq n$
is given by:
\[
PM\left(k\right)=W^{k}\cdot\prod_{i=1}^{k}r_{p_{i}}\cdot\prod_{i,j\leq k;i\leq j}sel_{p_{i},p_{j}}.
\]

The overall cost function we will attempt to minimize is thus the
sum of partial matches of all sizes, as follows:
\[
Cost_{ord}\left(O\right)=\sum_{k=1}^{n}\left(W^{k}\cdot\prod_{i=1}^{k}r_{p_{i}}\cdot\prod_{i,j\leq k;i\leq j}sel_{p_{i},p_{j}}\right).
\]

For the JQPG problem restricted to output left-deep trees only, we
will use the two-way join cost function $C\left(S,T\right)$ defined
in Section \ref{sub:Join-Query-Plan}. Let $L$ be a left-deep tree
and let $\left\{ i_{1},i_{2},\cdots,i_{n}\right\} $ be the order
in which input relations are to be joined according to $L$. Let $P_{k},1\leq k<n$
denote the result of joining the first $k$ tables by $L$ (that is,
$P_{1}=R_{i_{1}}$, $P_{2}=R_{i_{1}}\bowtie R_{i_{2}}$, etc.). In
addition, let $C_{1}=\left|R_{i_{1}}\right|\cdot f_{i_{1},i_{1}}$
be the cost of the initial selection from $R_{i_{1}}$. Then, the
cost of $L$ will be defined according to a left-deep join (LDJ) cost
function:
\[
Cost_{LDJ}\left(L\right)=C_{1}+\sum_{k=2}^{n}C\left(P_{k-1},R_{i_{k}}\right).
\]

We are now ready to formally prove the statement formulated in the
beginning of the section.

\newtheorem{thrm}{Theorem}
\begin{thrm}

Given a pure conjunctive pattern $P$, the problem of finding an order-based
evaluation plan for $P$ minimizing $Cost_{ord}$ is equivalent to
the Join Query Plan Generation problem for left-deep trees subject
to $Cost_{LDJ}$.

\end{thrm}

We will start by proving that $CPG\subseteq JQPG$. To that end, we
will present the corresponding reduction.

Given a pure conjunctive pattern $P$ defined as follows:

\[
\begin{array}{l}
\texttt{PATTERN AND}\left(T_{1}\: e_{1},\cdots T_{n}\: e_{n}\right)\\
\texttt{WHERE }\left(c_{1,1}\wedge c_{1,2}\wedge\cdots\wedge c_{n-1,n}\wedge c_{n,n}\right)\\
\texttt{WITHIN\: W,}
\end{array}
\]%

let $R_{1},\cdots,R_{n}$ be a set of relations such that each $R_{i}$
corresponds to an event type $T_{i}$. For each attribute of $T_{i}$,
including the timestamp, a matching column will be defined in $R_{i}$.
The cardinality of $R_{i}$ will be set to $W\cdot r_{i}$, and, for
each predicate $c_{i,j}$ with selectivity $sel_{i,j}$, an identical
predicate will be formed between the relations $R_{i}$ and $R_{j}$.
We will define the query corresponding to $P$ as follows:

\[
\begin{array}{l}
\texttt{SELECT *}\\
\texttt{FROM }R_{1},\cdots R_{n}\\
\texttt{WHERE }\left(c_{1,1} \texttt{ AND}\cdots\texttt{AND } c_{n,n}\right).\\
\end{array}
\]%

We will show that a solution to this instance of the JQPG problem
is also a solution to the initial CPG problem. Recall that a left-deep
JQPG solution $L$ minimizes the function $Cost_{LDJ}$. By opening
the recursion and substituting the parameters with those of the original
problem, we get:

\begin{align*}
&Cost_{LDJ}\left(L\right)=C_{1}+\sum_{k=2}^{n}C\left(P_{k-1},R_{i_{k}}\right)=\\
=&\left|R_{i_{1}}\right|\cdot f_{i_{1},i_{1}}+\sum_{k=2}^{n}\left(\prod_{j=1}^{k}\left|R_{i_{j}}\right|\cdot\prod_{j.l\leq k;j\leq l}f_{i_{j},i_{l}}\right)=\\
=&\sum_{k=1}^{n}\left(\prod_{j=1}^{k}\left(W\cdot r_{i_{j}}\right)\cdot\prod_{j.l\leq k-1;j\leq l}sel_{i_{j},i_{l}}\right)=Cost_{ord}\left(O\right).
\end{align*}%

Consequently, the solution that minimizes $Cost_{LDJ}$ also minimizes
$Cost_{ord}$, which completes the proof of the first direction of
the theorem.

We will now proceed to proving the second direction, i.e., $JQPG\subseteq CPG$.
Let $R_{1},\cdots,R_{n}$ be the relations to be joined with mutual
predicates $F_{i,j}$ of selectivities $f_{i,j}$. We will demonstrate
a decomposition of this problem to an instance of the CPG problem.

Let $T_{1},\cdots,T_{n}$ be primitive event types such that each
$T_{i}$ corresponds to a relation $R_{i}$. Let each instance of
$T_{i}$ have the attributes identical to the columns of $R_{i}$.
In addition, let $S$ be an input stream containing a primitive event
$t_{i,k}$ of type $T_{i}$ for each tuple $k$ of a relation $R_{i}$,
where $1\leq k\leq\left|R_{i}\right|$ is an index of a tuple in a
relation. Let the timestamp of each such event be defined as $t_{i,k}.ts=k$.
Define the time window as $W=max\left|R_{i}\right|$, where $\left|R_{i}\right|$
is the cardinality of $R_{i}$. Finally, let $r_{i}$, the arrival
rate of events of type $T_{i}$, be equal to $\frac{\left|R_{i}\right|}{W}$.

Now we are ready to define a CEP conjunctive pattern corresponding
to the join of $R_{1},\cdots,R_{n}$, as follows:

\[
\begin{array}{l}
\texttt{PATTERN AND}\left(T_{1}\: e_{1},\cdots T_{n}\: e_{n}\right)\\
\texttt{WHERE }\left(F_{1,1}\wedge F_{1,2}\wedge\cdots\wedge F_{n-1,n}\wedge F_{n,n}\right)\\
\texttt{WITHIN\: W.}
\end{array}
\]%

We will show that a solution to this instance of the CPG problem is
also a solution to the initial JQPG problem. Recall that such a solution
$O$ minimizes the function $Cost_{ord}$. By substituting the parameters
with those of the original problem, we get:

\begin{align*}
&Cost_{ord}\left(O\right)=\sum_{k=1}^{n}\left(W^{k}\cdot\frac{\prod_{i=1}^{k}\left|R_{p_{i}}\right|}{W^{k}}\cdot\prod_{i,j\leq k;i\leq j}f_{p_{i},p_{j}}\right)=\\
=&\sum_{k=1}^{n}\left(\left(\prod_{i=1}^{k-1}\left|R_{p_{i}}\right|\cdot\prod_{i,j\leq k-1;i\leq j}f_{p_{i},p_{j}}\right)\cdot\left|R_{p_{k}}\right|\cdot\prod_{i\leq k}f_{p_{i},p_{k}}\right).
\end{align*}

Here, the first sub-expression inside the summation is the intermediate
results size until the point when the $k^{th}$ relation is to be
joined, as follows from the definition above. The second sub-expression
is the cardinality of the relation $R_{p_{k}}$, and the third sub-expression
is the product of all predicates to be applied on tuples from $R_{p_{k}}$
upon joining it with $R_{p_{1}},\cdots R_{p_{k-1}}$. Therefore, the
resulting expression is identical to $Cost_{LDJ}$ (in particular,
the first element of the summation equals $C_{1}$). Consequently,
the solution that minimizes $Cost_{ord}$ also minimizes $Cost_{LDJ}$,
which completes the proof.$\blacksquare$

In \cite{CluetM95} the authors showed the problem of Join Query Plan
Generation for left-deep trees to be NP-complete for the general case of arbitrary query graphs. From this result
and from the above theorem we will deduce the following corollary.\newtheorem{corr}{Corollary}
\begin{corr}

The problem of finding an order-based evaluation plan for a general pure conjunctive
complex event pattern that minimizes $Cost_{ord}$ is NP-complete.

\end{corr}

\subsection{Tree-Based Evaluation}

\label{sub:Tree-Based-Evaluation}

In this section, we will extend Theorem 1 to tree-based evaluation
plans. This time we will consider the unrestricted
JQPG problem, allowed to return bushy trees. Similarly to the previous
section, we will start by defining the cost functions and then proceed
to the proof of the extended theorem.

We will define the cost model for evaluation trees in a manner similar
to Section \ref{sub:Order-Based-Evaluation}. We will estimate the
number of partial matches accumulated in each node of the evaluation
tree and sum them up to produce the cost function.

For a leaf node $l$ collecting events of type $T_{i}$, the expected
number of partial matches is equal to the number of events of type
$T_{i}$ arriving inside a time window:
\[
PM(l)=W\cdot r_{i}.
\]

To obtain an estimate for an internal node $in$, we multiply the
cost function values of its children by the total selectivity of the
predicates verified by this node:
\[
PM(in)=PM\left(in.left\right)\cdot PM\left(in.right\right)\cdot SEL_{LR}\left(in\right),
\]
where $SEL_{LR}$ is the selectivity of the predicates defined between
event types accepted at the left and the right sub-trees of node $in$,
or, more formally:
\[
SEL_{LR}\left(in\right)=\prod_{\begin{array}{c}
e_{i}\in in.ltree;e_{j}\in in.rtree\end{array}}sel_{i,j}.
\]

The total cost function on a tree $T$ is thus defined as follows:
\[
Cost_{tree}\left(T\right)=\sum_{N\in nodes\left(T\right)}PM\left(N\right).
\]

For bushy trees, we will extend the cost function defined in Section
\ref{sub:Order-Based-Evaluation}. The cost of a tree node $N$ will
be defined as follows:
\[
C\left(N\right)=\begin{cases}
\left|R_{i}\right| & N\: is\: a\: leaf\: representing\: R_{i}\\
\left|L\right|\cdot\left|R\right|\cdot f_{L,R} & N\: is\: an\: internal\: node\: representing\\
 & a\: sub-join\: L\bowtie R,
\end{cases}
\]
with the bushy join (BJ) cost function defined as follows:
\[
Cost_{BJ}\left(T\right)=\sum_{N\in nodes\left(T\right)}C\left(N\right).
\]

We will now extend Theorem 1 to tree-based plans.

\begin{thrm}

Given a pure conjunctive pattern $P$, the problem of finding a tree-based
evaluation plan for $P$ minimizing $Cost_{tree}$ is equivalent to
the Join Query Plan Generation problem subject to $Cost_{BJ}$.

\end{thrm}

To prove the theorem, we decompose each of the tree cost functions
$Cost_{tree},Cost_{BJ}$ defined above for the CPG and the JQPG problems
into two components, separately calculating the cost of the leaves
and the internal nodes:

\begin{align*}
&Cost_{tree}^{l}\left(T\right)=\sum_{N\in leaves\left(T\right)}PM\left(N\right)\\
&Cost_{tree}^{in}\left(T\right)=\sum_{N\in in\_nodes\left(T\right)}PM\left(N\right)\\
&Cost_{BJ}^{l}\left(T\right)=\sum_{N\in leaves\left(T\right)}C\left(N\right)\\
&Cost_{BJ}^{in}\left(T\right)=\sum_{N\in\ in\_nodes\left(T\right)}C\left(N\right).
\end{align*}

Obviously, the following equalities hold:

\begin{align*}
&Cost_{tree}\left(T\right)=Cost_{tree}^{l}\left(T\right)+Cost_{tree}^{in}\left(T\right)\\
&Cost_{BJ}\left(T\right)=Cost_{BJ}^{l}\left(T\right)+Cost_{BJ}^{in}\left(T\right).
\end{align*}

Thus, it is sufficient to prove that

\begin{align*}
&Cost_{tree}^{l}\left(T\right)=Cost_{BJ}^{l}\left(T\right)\\
&Cost_{tree}^{in}\left(T\right)=Cost_{BJ}^{in}\left(T\right).
\end{align*}

for every $T$. From here it will follow that the solution minimizing $Cost_{tree}$
will also minimize $Cost_{BJ}$ and vice versa.

Applying either direction of the reduction from Theorem 1, we observe
the following for the first pair of functions:

\begin{multline*}
Cost_{tree}^{l}\left(T\right)=\sum_{N\in leaves\left(T\right)}PM\left(N\right)=\sum_{i=1}^{n}W\cdot r_{i}=\\
=\sum_{i=1}^{n}\left|R_{i}\right|=\sum_{N\in leaves\left(T\right)}C\left(N\right)=Cost_{BJ}^{l}\left(T\right).
\end{multline*}

Similarly, for the second pair of functions:

\begin{multline*}
Cost_{tree}^{in}\left(T\right)=\sum_{N\in in\_nodes\left(T\right)}PM\left(N\right)=\\
=\sum_{N\in in\_nodes\left(T\right)}PM\left(N.left\right)\cdot PM\left(N.right\right)\cdot SEL_{LR}\left(N\right).
\end{multline*}

Opening the recursion, we get:

\begin{multline*}
Cost_{tree}^{in}\left(T\right)=
\sum_{N\in in\_nodes\left(T\right)}\left(\prod_{m\in leaves\left(N\right)}W\cdot r_{m}\cdot\prod_{i,j\in leaves\left(N\right)}sel_{i,j}\right).
\end{multline*}

By applying an identical transformation on $Cost_{JB}^{in}$ we obtain
the following:

\begin{multline*}
Cost_{BJ}^{in}\left(T\right)=
\sum_{N\in in\_nodes\left(T\right)}\left(\prod_{m\in leaves\left(N\right)}\left|R_{m}\right|\cdot\prod_{i,j\in leaves\left(N\right)}f_{i,j}\right).
\end{multline*}

After substituting $r_{m}=\frac{\left|R_{m}\right|}{W}$ and $sel_{p_{i},p_{j}}=f_{p_{i},p_{j}}$,
the two expressions are identical, which completes the proof.$\blacksquare$

The CPG-JQPG reduction that we will use for tree-based evaluation
is the one demonstrated in Theorem 1 for order-based evaluation.

By Theorem 2 and the generalization of the result in \cite{CluetM95},
we derive the following corollary.

\begin{corr}

The problem of finding a tree-based evaluation plan for a general pure conjunctive
complex event pattern that minimizes $Cost_{tree}$ is NP-complete.

\end{corr}

\subsection{Join Query Types}

\label{sub:Discussion}

As Corollaries 1 and 2 imply, no efficient algorithm can be devised
to optimally solve CPG for a general conjunctive pattern unless $P=NP$.
However, better complexity results may be available under certain
assumptions regarding the pattern structure. Numerous works considered
the JQPG problem for restricted \textit{query types}, that is, specific
topologies of the query graph defining the inter-relation conditions.
Examples of such topologies include clique, tree, and star.

It was shown in \cite{Ibaraki84,KrishnamurthyBZ86} that an optimal
plan can be computed in polynomial time for left-deep trees and queries
forming an acyclic graph (i.e., tree queries), provided that the cost
function has the ASI (adjacent sequence interchange) property \cite{MonmaS79}.
The left-deep tree cost function $Cost_{LDJ}$ has this property \cite{CluetM95},
making the result applicable for our scenario. A polynomial algorithm without
the ASI requirement was proposed for bushy tree plans for chain queries
\cite{Orlowski14}. From Theorems 1 and 2 we can conclude that,
for conjunctive patterns only, CPG$\in$P under the above constraints.

However, these results only hold when the plans produced by a query
optimizer are not allowed to contain cross products \cite{CluetM95,Orlowski14}.
While this limitation is well-known in relational optimization \cite{VanceM96},
it is not employed by the existing CPG methods \cite{AkdereMCT08,KolchinskySS15,MeiM09,Schultz-MollerMP09}.
Moreover, it was shown that when cross products are omitted, cheaper plans might be missed
\cite{OnoL90}. Thus, even when an exact polynomial algorithm
is applicable to CPG, it is inferior to native algorithms in terms
of the considered search space and can only be viewed as a heuristic.
In that sense, it is similar to the  greedy and randomized approaches \cite{SteinbrunnMK97,Swami89}.

Other optimizations utilizing the knowledge of the query type were proposed.
For example, the optimal bushy plan was empirically shown to be identical
to the optimal left-deep plan for star queries and, in many cases,
for grid queries \cite{SteinbrunnMK97}. This observation allows us to
utilize a cheaper left-deep algorithm for the above query types without
compromising the quality of the resulting plan.

With the introduction of additional pattern types (Section \ref{sec:General-Pattern-Types})
and event selection strategies (Section \ref{sub:Event-Consumption-Policies}),
new query graph topologies might be identified and type-specific efficient
algorithms designed. This topic is beyond the scope of this paper
and is a subject for future work.

Although not used directly by the JQPG algorithms, the order-based
CPG cost functions $Cost_{ord}$ and $Cost_{ord}^{lat}$ (that we
will introduce in Section \ref{sub:Pattern-Detection-Latency}) also
have the ASI property. We formally prove this statement in the Appendix \ref{sec:ASI-Property-of}.

\section{JQPG for General Pattern Types}

\label{sec:General-Pattern-Types}

The reduction from CPG to JQPG presented above only applies to pure
conjunctive patterns. However, the patterns encountered in real-world
scenarios are much more diverse. To complete the solution, we have
to consider simple patterns containing SEQ, OR, NOT and KL operators.
We also have to address nested patterns.

This section describes how a pattern of each of the aforementioned
types can be represented and detected as either a pure conjunctive
pattern or their union. To that end, we will utilize some of the ideas
from \cite{KolchinskySK17Long}. Note that the transformations presented
below are only applied for the purpose of plan generation, that is,
no actual conversion takes place during execution on a data stream.

\subsection{Sequence patterns}

We observe that a sequence pattern is merely a conjunctive pattern
with additional temporal constraints, i.e., predicates on the values
of the timestamp attribute. Thus, a general pure sequence pattern
of the form

\[
\begin{array}{l}
PATTERN\: SEQ\left(T_{1}\: e_{1},T_{2}\: e_{2},\cdots,T_{n}\: e_{n}\right)\\
WHERE\:(c_{1,1}\wedge c_{1,2}\wedge\cdots\wedge c_{n,n-1}\wedge c_{n,n})\\
WITHIN\: W
\end{array}
\]%
can be rewritten in the following way without any change in the semantics:

\[
\begin{array}{l}
PATTERN\: AND\left(T_{1}\: e_{1},T_{2}\: e_{2},\cdots,T_{n}\: e_{n}\right)\\
WHERE\:(c_{1,1}\wedge\cdots\wedge c_{n,n}\wedge\\
\qquad\qquad\wedge\left(e_{1}.ts<e_{2}.ts\right)\wedge\cdots\wedge\left(e_{n-1}.ts<e_{n}.ts\right))\\
WITHIN\: W.
\end{array}
\]%

An instance of the sequence pattern is thus reduced from CPG to JQPG
similarly to a conjunctive pattern, with the \textit{timestamp} column
added to each relation $R_{i}$ representing an event type $T_{i}$,
and constraints on the values of this column introduced into the query
representation.

We will now formally prove the correctness of the above construction.

\begin{thrm}

Let $S$ be a pure sequence pattern specified by primitive event types
$T_{1},\cdots,T_{n}$, a Boolean predicate of constraints $P_{S}=\bigwedge_{1\leq i,j\leq n}c_{i,j}$,
and a time window $W$. Additionally, let $C$ be a pure conjunctive
pattern specified by $T_{1},\cdots,T_{n}$, a time window $W'=W$,
and a predicate $P_{C}=P_{S}\wedge P_{O}$, where $P_{O}=\bigwedge_{1\leq i<n}\left(e_{i}.ts<e_{i+1}.ts\right)$.
Then, $S$ is equivalent to $C$, i.e., both patterns specify the
same set of matches.

\end{thrm}

We will prove this theorem by double inclusion.

$S\subseteq C$: Let $M=\left\{ e_{1},\cdots,e_{n}\right\} $ be a
match for $S$. Then, by definition, $M$ satisfies $P_{S}$. In addition,
since $S$ is a sequence pattern, events in $M$ follow the temporal
order $e_{1}.ts<e_{2}.ts<\cdots<e_{n}.ts$. Hence, $M$ also satisfies
$P_{O}$, and thus $P_{C}$ as well, i.e., $M$ is a match for $C$.

$C\subseteq S$: Let $M=\left\{ e_{1},\cdots,e_{n}\right\} $ be a
match for $C$. Since $M$ satisfies $P_{O}$, its events satisfy
the ordering constraints $e_{1}.ts<\cdots<e_{n}.ts$. By definition
of a sequence pattern, $\left\{ e_{1},\cdots,e_{n}\right\} $ form
a valid sequence within the time window $W$. Additionally, $M$ satisfies
$P_{S}$. Hence, by definition, $M$ is a match for $S$.$\blacksquare$

\subsection{Kleene closure patterns}

In a pattern with an event type
$T_{i}$ under a KL operator, any subset of events of $T_{i}$ within
the time window can participate in a match. During plan generation,
we are interested in modeling this behavior in a way comprehensible by
a JQPG algorithm, that is, using an equivalent pattern without Kleene
closure. To that end, we introduce a new type $T_{i}'$ to represent
all event subsets accepted by $KL\left(T_{i}\right)$, that is, the
power set of events of $T_{i}$. A set of $k$ events of type $T_{i}$
will be said to contain $2^{k}$ ``events'' of type $T_{i}'$, one
for each subset of the original $k$ events. The new pattern is constructed
by replacing $KL\left(T_{i}\right)$ with $T_{i}'$. Since a time
window of size $W$ contains $2^{r_{i}\cdot W}$ subsets of $T_{i}$
(where $r_{i}$ is the arrival rate of $T_{i}$), the arrival rate
$r{}_{i}'$ of $T_{i}'$ is set to $\frac{2^{r_{i}\cdot W}}{W}$.
The predicate selectivities remain unchanged.

For example, given the following pattern with the arrival rate of
5 events per second for each event type:

\[
\begin{array}{l}
\texttt{PATTERN AND(A a,KL(B b),C c)}\\
\texttt{WHERE (\textit{true})  WITHIN 10 seconds,}
\end{array}
\]%

the pattern to be utilized for plan generation will be:

\[
\begin{array}{l}
\texttt{PATTERN AND(A a,B' b,C c)}\\
\texttt{WHERE (\textit{true})  WITHIN 10 seconds.}
\end{array}
\]%

The arrival rate of $B'$ will be calculated as $r_{B}'=\frac{2^{r_{B}\cdot W}}{W}=\frac{1}{10}\cdot2^{50}$.
A plan generation algorithm will then be invoked on the new pattern.
Due to an extremely high arrival rate of $B'$, its processing will
likely be postponed to the latest step in the plan, which is also
the desired strategy for the original pattern in this case. $B'$
will then be replaced with $B$ in the resulting plan, and the missing
Kleene closure operator will be added in the respective stage (by
modifying an edge type for a NFA \cite{KolchinskySK17Long} or a node
type for a tree \cite{MeiM09}), thus producing a valid plan for detecting
the original pattern.
We will now formally prove the correctness of the above construction.

\begin{thrm}

Let $K$ be a conjunctive pattern specified by primitive event types
$T_{1},\cdots,T_{i},\cdots,T_{n}$, a Boolean predicate of constraints
$P_{K}=\bigwedge_{1\leq i,j\leq n}c_{i,j}$, and a time window $W$.
In addition, let $K$ contain a Kleene closure operator applied on
an event type $T_{i}$. Let a new type $T_{i}'$ represent the power
set of events of the type $T_{i}$, i.e., for each set of events $\left\{ e_{i}^{1},e_{i}^{2},\cdots,e_{i}^{k}\right\} $
of type $T_{i}$ within the time window an event of type $T{}_{i}'$
is created, containing each of $e_{i}^{j}$ as an attribute. Let $C$
be a pure conjunctive pattern specified by $T_{1},\cdots,T_{i}',\cdots,T_{n}$,
a predicate $P{}_{C}=P_{K}$, and a time window $W'=W$. Then, $K$
is equivalent to $C$, i.e., both patterns specify the same set of
matches.

\end{thrm}

We will show that for an arbitrary input stream both patterns will
produce identical sets of pattern matches. Let $S=\left\{ s_{1},\cdots,s_{m}\right\} $
be an input stream. W.l.o.g., assume that each event in $S$ belongs
to one of the types $T_{1},\cdots,T_{n}$ and that all events are
within the time window $W$. Additionally, let $S'\subset S$ denote
the set of all events of type $T_{i}$ in $S$. Then, while monitoring
the pattern $K$, the system will create $2^{\left|S'\right|}-1$
partial matches for each unique combination of events of types $T_{1},\cdots,T_{i-1},T_{i+1},\cdots,T_{n}$.

Now, let $S''\subset S$ denote the set of all events of type $T{}_{i}'$
in $S$. While monitoring $C$, for each combination of events of
types $T_{1},\cdots,T_{i-1},T_{i+1},\cdots,T_{n}$ the detecting framework
will create $\left|S''\right|$ partial matches, one for each primitive
event in $S''$. For each non-empty subset of $S'$, $S''$ contains
an event corresponding to this subset, and vice versa. As the constraints
on both patterns are identical, as well as the time window, the resulting
full matches will also be the same.$\blacksquare$

\begin{corr}

Theorem 4 holds also for sequence patterns.

\end{corr}

The correctness of this corollary follows from the transitivity of
conversions in Theorems 3 and 4.

\begin{corr}

Theorem 4 and Corollary 3 hold for an arbitrary number of non-nested
Kleene closure operators in a pattern.

\end{corr}

The proof of this corollary is by iteratively applying Theorem 4 or
Corollary 3 on each Kleene closure operator.

\subsection{Negation patterns}

Patterns with a negated event will not be rewritten. Instead, we will
introduce a negation-aware evaluation plan creation strategy. First,
a plan will be generated for a positive part of a pattern as described
above. Then, a check for the appearance of a negated event will be
added at the earliest point possible, when all positive events it
depends on are already received. This construction process will be
implemented by augmenting a plan with a transition to the rejecting
state for a NFA \cite{KolchinskySK17Long} or with a NSEQ node for
a ZStream tree \cite{MeiM09}. For example, given a pattern $SEQ\left(A,NOT\left(B\right),C,D\right)$,
the existence of a matching $B$ in the stream will be tested immediately
after the latest of $A$ and $C$ have been accepted. Since both Lazy
NFA and ZStream incorporate event buffering, this technique is feasible
and easily applicable.

\subsection{Nested patterns}

Patterns of this type can contain an unlimited number of n-ary operators.
After transforming SEQ to AND as shown above, we are left with only
two such operator types, AND and OR. Given a nested pattern, we convert
the pattern formula to DNF form, that is, an equivalent nested disjunctive
pattern containing a list of simple conjunctive patterns is produced.
Then, a separate evaluation plan is created for each conjunctive subpattern,
and their detection proceeds independently. The returned result is
the union of all subpattern matches.

Note that applying the DNF transformation can cause some expressions
to appear in multiple subpatterns. For example, a nested pattern of
the form $AND\left(A,B,OR\left(C,D\right)\right)$ will be converted
to a disjunction of conjunctive patterns $AND\left(A,B,C\right)$
and $AND\left(A,B,D\right)$. As a result, redundant computations
will be performed by automata or trees corresponding to different
subpatterns (comparing $A$'s to $B$'s in our example). This problem
can be solved by applying known multi-query techniques for shared
subexpression processing, such as those described in \cite{DemersGHRW06,LiuRGGWAM11,RayLR16,RayRLGWA13,ZhangVDH17}.

\section{Adapting JQPG Algorithms to \newline Complex Event Processing}

\label{sec:Adapting-Join-Query}

The theoretical results from previous sections imply that any existing technique 
for determining a close-to-optimal execution plan for a join query can be adapted and used in CEP applications.
However, many challenges arise when attempting to perform this transformation
procedure in practice. First, despite the benefits of the cost function
introduced in Section \ref{sub:CEP-Plan-Generation}, simply counting
the partial matches is not always sufficient. Additional performance
metrics are often essential, such as the average response time. Second,
complex event specification languages contain various constructs not
present in traditional databases, such as event selection strategies.
Third, the arrival rates of event types and the predicate selectivities
are rarely obtained in advance and can change rapidly over time. A
solution must be devised to measure the desired statistics on-the-fly
and adapt the evaluation plan accordingly.

In this section, we show how detection latency and event selection
strategies can be incorporated into existing JQPG algorithms. We also
address the problem of adapting to dynamic changes in the input stream.

\subsection{Pattern Detection Latency}

\label{sub:Pattern-Detection-Latency}

Latency is commonly defined as a time difference between the arrival
of the last primitive event comprising a full pattern match and the
time of reporting this match by a CEP system. As many existing applications
involve strong real-time requirements, pattern detection latency has
become an important optimization goal for such systems. Unfortunately,
in most cases it is impossible to simultaneously achieve maximal throughput
and minimal latency, and trade-offs between the two are widely studied
in the context of complex event processing \cite{AkdereMCT08,YiLW16}.

Detection schemes utilizing out-of-order evaluation, like those discussed
in this paper, often suffer from increased latency as compared to
more na\"{i}ve approaches. The main reason is that, when an execution
plan is optimized for maximal throughput, the last event in the pattern
may not be the last event in the plan. After this event is accepted,
the evaluation mechanism still needs to walk through the remaining
part of the plan, resulting in late detection of the full match.

Algorithms adopted from JQPG do not naturally support latency. However,
since they are generally independent of the cost model, this problem
can be solved by providing an appropriate cost function. In addition
to functions presented in Sections \ref{sub:Order-Based-Evaluation}
and \ref{sub:Tree-Based-Evaluation}, which we will refer to as $Cost_{ord}^{trpt}$
and $Cost_{tree}^{trpt}$, a new pair of functions, $Cost_{ord}^{lat}$
and $Cost_{tree}^{lat}$, will reflect the expected latency of
a plan. To combine the functions, many existing multi-objective query
optimization techniques can be used, e.g., pareto optimal plan calculation
\cite{AkdereMCT08} or parametric methods \cite{TrummerK16}. Systems
with limited computational resources may utilize simpler and less
expensive solutions, such as defining the total cost function as a
weighted sum of its two components:
\[
Cost\left(Plan\right)=Cost^{trpt}\left(Plan\right)+\alpha\cdot Cost^{lat}\left(Plan\right),
\]
where $\alpha$ is a user-defined parameter adjusted to fit the required
throughput-latency trade-off. This latter model was used during our
experiments (Section \ref{sec:Experimental-Evaluation}).

We will now formally define the latency cost functions. For a sequence
pattern, let $T_{n}$ denote the last event type in the order induced
by the pattern. Then, for an order-based plan $O$, let $Succ_{O}\left(T_{n}\right)$
denote the event types succeeding $T_{n}$ in $O$. Following the
arrival of an event of type $T_{n}$, in the worst case we need to
examine all locally buffered events of types in $Succ_{O}\left(T_{n}\right)$.
As defined in Section \ref{sub:Order-Based-Evaluation}, there are
$W\cdot r_{i}$ such events of type $T_{i}$, hence:
\[
Cost_{ord}^{lat}\left(O\right)=\sum_{T_{i}\in Succ_{O}\left(T_{n}\right)}W\cdot r_{i}.
\]

Similarly, for a tree-based plan $T$, let $Anc_{T}\left(T_{n}\right)$
denote all ancestor nodes of the leaf corresponding to $T_{n}$ in
$T$, i.e., nodes located on a path from $T_{n}$ to the root (excluding
the root). Let us examine the traversal along this path. When an internal
node $N$ with two children $L$ and $R$ receives a partial match
from, say, the child $L$, it compares this match to all partial matches
currently buffered on $R$. Thus, the worst-case detection latency
of a sequence pattern ending with $T_{n}$ is proportional to the
number of partial matches buffered on the siblings of the nodes in
$Anc_{T}\left(T_{n}\right)$. More formally, let $sibling\left(N\right)$
denote the other child of the parent of $N$ (for the root this function
will be undefined). Then,
\[
Cost_{tree}^{lat}\left(T\right)=\sum_{N\in Anc_{T}\left(T_{n}\right)}PM\left(sibling\left(N\right)\right).
\]

For a conjunctive pattern, estimating the detection latency is a more
difficult problem, as the last arriving event is not known in advance.
One possible approach is to introduce a new system component, called
the output profiler. The output profiler examines the full matches
reported as output and records the most frequent temporal orders in
which primitive events appear. Then, as enough information is collected,
the latency function may be defined as in the previous case, subject
to the event arrival order with the highest probability of appearance.

Finally, for a disjunctive pattern, we define the latency cost function
as the maximum over the disjunction operands. This definitions applies
also for arbitrary nested patterns.

\subsection{Event Selection Strategies}

\label{sub:Event-Consumption-Policies}

In addition to event types, operators and predicates, CEP patterns
are further defined using the \textit{event selection strategies}
\cite{AgrawalDGI2008,CugolaM12,EtzionN10} . An event selection strategy
specifies how events are selected from an input stream for partial
matches. In this section, we discuss four existing strategies and
show how a reduction from JQPG to CPG can support them.

Until now, we have implicitly assumed the \textit{skip-till-any-match} selection
strategy \cite{AgrawalDGI2008}, which permits a primitive event to
participate in an unlimited number of matches. This strategy is the
most flexible, as it allows all possible combinations of events comprising
a match to be detected. However, some streaming applications do not
require such functionality. Thus, additional strategies were defined,
restricting the participation of an event in a match.

The \textit{skip-till-next-match} selection strategy \cite{AgrawalDGI2008}
limits a primitive event to appear in no more than a single full match.
This is enforced by ``consuming'' events already assigned to a match.
While this strategy prevents some matches from being discovered, it
also considerably simplifies the detection process. In a CEP system
operating under the skip-till-any-match policy, our cost model will
no longer provide a correct estimate for a number of partial matches,
which would lead to arbitrarily inefficient evaluation plans. However,
since most JQPG algorithms do not depend on a specific cost function,
we can solve this issue by replacing $Cost_{order}$ and $Cost_{tree}$
with newly devised models.

Let us examine the number of partial matches in an order-based setting
under the skip-till-next-match strategy. We will denote by $m\left[k\right]$
the number of matches of size $k$ expected to exist simultaneously
in a time window. Obviously, $m\left[1\right]=W\cdot r_{p_{1}}$,
where $T_{p_{1}}$ is the first event type in the selected evaluation
order. For the estimate of $m\left[2\right]$, there are two possibilities.
If $r_{p_{1}}>r_{p_{2}}$, there will not be enough instances of $T_{p_{2}}$
to match all existing instances of $T_{p_{1}}$, and some of the existing
matches of size 1 will never be extended. Hence, $m\left[2\right]=W\cdot r_{p_{2}}$
in this case. Otherwise, as an existing partial match cannot be extended
by more than a single event of type $T_{p_{2}}$, $m\left[1\right]$
will be equal to $m\left[2\right]$. In addition, if a mutual condition
exists between $T_{p_{1}}$and $T_{p_{2}}$, the resulting expression
has to be multiplied by $sel_{p_{1},p_{2}}$.

By extending this reasoning to an arbitrary partial match, we obtain
the following expression:
\[
m\left[k\right]=W\cdot min\left(r_{p_{1}},r_{p_{2}},\cdots,r_{p_{k}}\right)\cdot\prod_{i,j\leq k;i\leq j}sel_{p_{i},p_{j}};\:1\leq k\leq n.
\]

And the new cost function for order-based CPG is
\[
Cost_{ord}^{next}\left(O\right)=\sum_{k=1}^{n}\left(W\cdot m\left[k\right]\right).
\]

Using similar observations, we extend the above result for the tree-based
model:

\begin{align*}
&PM(n)=W\cdot min_{T_{i}\in subtree\left(n\right)}\left(r_{i}\right)\cdot\prod_{T_{i},T_{j}\in subtree\left(n\right)}sel_{i,j};\\
&Cost_{tree}^{next}\left(T\right)=\sum_{n\in nodes\left(T\right)}PM\left(n\right).
\end{align*}%

The two remaining selection strategies, \textit{strict contiguity}
and \textit{partition contiguity} \cite{AgrawalDGI2008}, define further
restrictions on the appearance of events in a match. The strict contiguity
requirement forces the selected events to be contiguous in the input
stream, i.e., it allows no other events to appear in between. The
partition contiguity strategy is a slight relaxation of the above.
It partitions the input stream according to some condition and only
requires the events located in the same partition to be contiguous.

To support JQPG-based solutions for CPG under strict or partition
contiguity, we will explicitly model the constraints imposed by the
above strategies. In addition, the cost model presented earlier for
skip-till-next-match will be used for both selection strategies.

To express strict contiguity, we will augment each primitive event
with a new attribute reflecting its unique serial number in the stream.
Then, we will add a new condition for each pair of potentially neighboring
events, requiring the numbers to be adjacent.

For partition contiguity, the new attribute will represent an inner,
per-partition order rather than a global one. Unless the partitioning
condition is very costly to evaluate (which is rarely the case), this
transformation can be efficiently and transparently applied on the
input stream. The new contiguity condition will first compare the
partition IDs of the two events, and only verify their serial numbers
if the IDs match.  We assume that the value
distribution across the partitions remains unchanged. Otherwise, the
evaluation plan is to be generated on a per-partition basis. Techniques
incorporating per-partition plans are beyond the scope of this paper
and are a subject for our future research.

\subsection{Adaptive Complex Event Processing}

\label{sub:Adaptive-Complex-Event}

As their definition implies, JQPG algorithms can only be used when
event arrival rates and predicate selectivities are given in advance.
However, in real-life scenarios this a priori knowledge is rarely
available. Moreover, the data characteristics are subject to frequent
on-the-fly fluctuations. To ensure efficient operation, a CEP engine
must continuously estimate the current statistic values and, when
a significant deviation is detected, adapt itself by recalculating
the affected evaluation plans. Developing efficient adaptive mechanisms
is considered a hard problem and a hot research topic in several fields
\cite{BabuMMNW04,DeshpandeIR07,KolchinskySS15,MeiM09}.

Due to the considerable generality, importance, and complexity of
adaptive complex event processing, we devote a separate paper \cite{ArxivAdaptivity} 
to the discussion of this problem. In it, we propose a novel
adaptivity mechanism and study it theoretically and empirically in
conjunction with a JQPG-based evaluation plan generator.

\section{Experimental Evaluation}

\label{sec:Experimental-Evaluation}

In this section, we present our experimental study on real-world data.
Our main goal was to compare some of the well-known JQPG algorithms,
adapted for CPG as described above, to the currently used methods
developed directly for CPG. The results demonstrate the superiority
of the former in terms of quality and scalability of the generated
plans.

In the following section we describe the algorithms compared during
the study. Then we present the experimental setup, followed by the
obtained results.

\subsection{CPG and JQPG Algorithms}

\label{sub:Evaluation-Algorithms}

We implemented 5 order-based and 3 tree-based CPG algorithms. Out
of those, 3 order-based and 2 tree-based algorithms are JQPG methods
adapted to the CEP domain. Our main goal is to evaluate those algorithms
against the rest, which are native CPG techniques.
The order-based plan generation algorithms included the following:

\textbullet{} Trivial order (TRIVIAL) - the evaluation plan is set
to the initial order of the sequence pattern. This strategy is used
in various CEP engines based on NFAs, such as SASE \cite{WuDR06} and
Cayuga \cite{DemersJB07}.

\textbullet{} Event frequency order (EFREQ) - the events are processed
by the ascending order of their arrival frequencies. This is the algorithm
of choice for frameworks such as PB-CED \cite{AkdereMCT08} and the
Lazy NFA \cite{KolchinskySS15}.

\textbullet{} Greedy cost-based algorithm (GREEDY) \cite{Swami89}
- this greedy heuristic algorithm for JQPG proceeds by selecting at
each step the relation which minimizes the value of the cost function.
Here and below, unless otherwise stated, we will use cost functions
minimizing the intermediate results size (Sections \ref{sub:Order-Based-Evaluation}
and \ref{sub:Tree-Based-Evaluation}).

\textbullet{} Iterative improvement algorithm (II-RANDOM / II-GREEDY)
- a local search JQPG algorithm, starting from some initial execution
plan and attempting a set of moves to improve the cost function, until
a local minimum is reached. In this study, we experimented with two
variations of this algorithm, presented in \cite{Swami89}. The first,
denoted as II-RANDOM, starts from a random order. The second, denoted
as II-GREEDY, first applies a greedy algorithm to create an initial
state. In both cases, the functions used to traverse between states
are \textit{swap} (the positions of two event types in a plan are
swapped) and \textit{cycle} (the positions of three event types are
shifted).

\textbullet{} Dynamic programming algorithm for left-deep trees (DP-LD)
- first presented in \cite{SelingerACLP79}, this exponential-time
algorithm utilizes dynamic programming to produce a provably optimal
execution plan. The result is limited to a left-deep tree topology.

For the tree-based plan generation algorithms, the following were
used:

\textbullet{} ZStream plan generation algorithm (ZSTREAM) - creates
an evaluation tree by iterating over all possible tree topologies
for a given sequence of leaves \cite{MeiM09}.

\textbullet{} ZStream with greedy cost-based ordering (ZSTREAM-ORD)
- as was demonstrated in Section \ref{sub:Tree-based-Evaluation-Mechanisms},
the limitation of the ZStream algorithm is in its inability to modify
the order of tree leaves. This algorithm attempts to utilize an order-based
JQPG method to overcome this drawback. It operates by first executing
GREEDY on the leaves of the tree to produce a 'good' ordering, then
applying ZSTREAM on the resulting list.

\textbullet{} Dynamic programming algorithm for bushy trees (DP-B)
\cite{SelingerACLP79} - same as DP-LD, but without the topology restriction.

\subsection{Experimental Setup}

\label{sub:Experimental-Setup}

The data used during the experiments was taken from the NASDAQ stock
market historical records \cite{EODData}. In this dataset, each record
represents a single update to the price of a stock, spanning a 1-year
period and covering over 2100 stock identifiers with prices periodically
updated. Our input stream contained 80,509,033 primitive events, each
consisting of a stock identifier, a timestamp, and a current price.
For each identifier, a separate event type was defined. In addition,
we augmented the event format to include the difference between the
current and the previous price of each stock. The differences were
calculated during the preprocessing stage.

To compare a set of plan generation algorithms, we need to use them
to create a set of evaluation plans for the same pattern and apply
the resulting plans on the input data stream using a CEP platform
of choice. To that end, we implemented two evaluation mechanisms discussed
in this paper, the out-of-order lazy NFA \cite{KolchinskySS15} and
the instance-based tree model based on ZStream \cite{MeiM09} as presented
in Section \ref{sub:Tree-based-Evaluation-Mechanisms}. The former
was then used to evaluate plans created by each order-based CPG or
JQPG algorithm on the patterns generated as described below. The latter
was similarly used for comparing tree-based plans.

The majority of the experiments were performed separately on 5 sets
of patterns: (1)pure sequences; (2)sequences with a negated event
(marked as 'negation' patterns in the graphs below); (3)conjunctions;
(4)sequences containing an event under KL operator (marked as 'Kleene
closure' patterns); (5)composite patterns, consisting of a disjunction
of three sequences (marked as 'disjunction' patterns). Each set contained
500 patterns with the sizes (numbers of the participating events)
ranging from 3 to 7, 100 patterns for each value. The pattern time
window was set to 20 minutes.

The pattern structure was motivated by the problem of monitoring the
relative changes in stock prices. Each pattern included a number of
predicates, roughly equal to half the size of a pattern, comparing
the \textit{difference} attributes of two of the involved event types.
For example, one pattern of size 3 from the set of conjunction patterns
was defined as follows:

\[
\begin{array}{l}
PATTERN\: AND(\\
\qquad MSFT \textunderscore Stock\: m, GOOG \textunderscore Stock\: g, INTC \textunderscore Stock\: i)\\
WHERE\:(m.difference < g.difference)\\
WITHIN\: 20\: minutes.
\end{array}
\]

The intention of this particular pattern is to examine the shift in
the value of Intel's stock in situations where Google's stock price
change is higher than Microsoft's.

All arrival rates and predicate selectivities were calculated during
the preprocessing stage. The measured arrival rates varied between
1 and 45 events per second, and the selectivities ranged from 0.002 to 0.88. As discussed
in Section \ref{sub:Adaptive-Complex-Event}, in most real-life scenarios
these statistics are not available in advance and may fluctuate frequently
and significantly during runtime. We experimentally study the impact
of these issues in a separate paper \cite{ArxivAdaptivity}.

We selected throughput and memory consumption as our performance metrics
for this study. Throughput was defined as the number of primitive
events processed per second during pattern detection using the selected
plan. To estimate the memory consumption, we measured the peak memory
required by the system during evaluation. The metrics were acquired
separately for each pattern, and the presented results were then calculated
by taking the average.

All models and algorithms under examination were implemented in Java.
The experiments were run on a machine with 2.20 Ghz CPU and 16.0 GB
RAM and took about 1.5 months to complete.

\subsection{Experimental Results}

\label{sub:Experimental-Results}

\begin{figure}
	\centering
	\subfloat[]{\includegraphics[width=.6\linewidth]{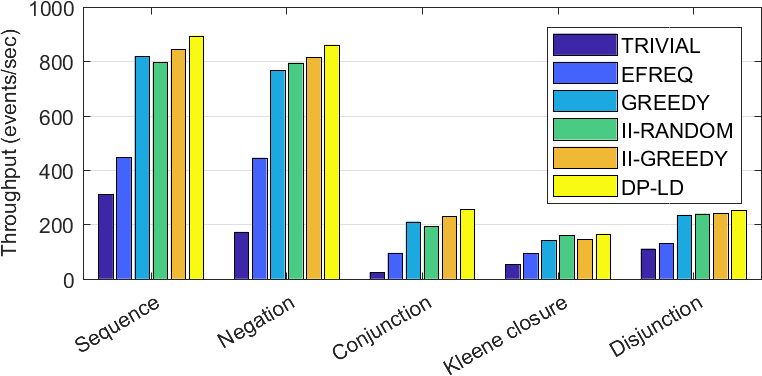}\label{fig:throughput-order}}\quad
	\subfloat[]{\includegraphics[width=.6\linewidth]{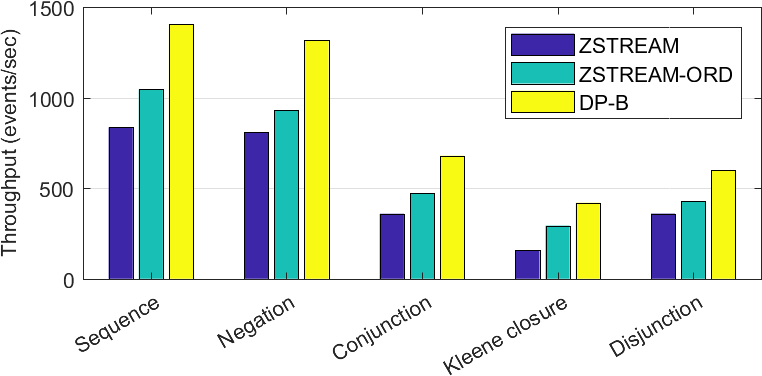}\label{fig:throughput-tree}}
    \caption{Throughput for different pattern types (higher is better): \protect\subref{fig:throughput-order} order-based methods; \protect\subref{fig:throughput-tree} tree-based methods.}
	\label{fig:throughput-by-type}
\end{figure}

\begin{figure}
	\centering
	\subfloat[]{\includegraphics[width=.6\linewidth]{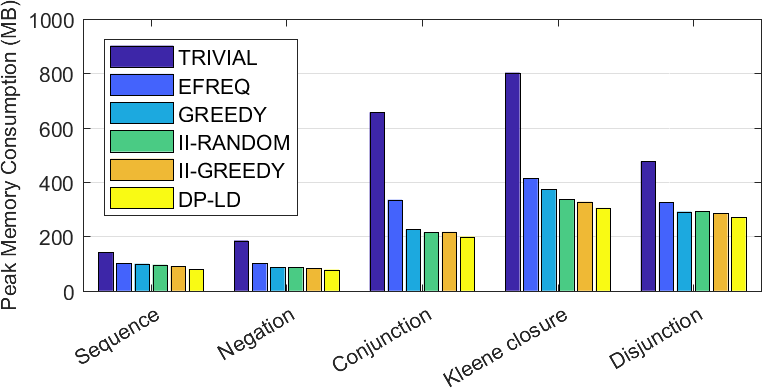}\label{fig:memory-order}}\quad
	\subfloat[]{\includegraphics[width=.6\linewidth]{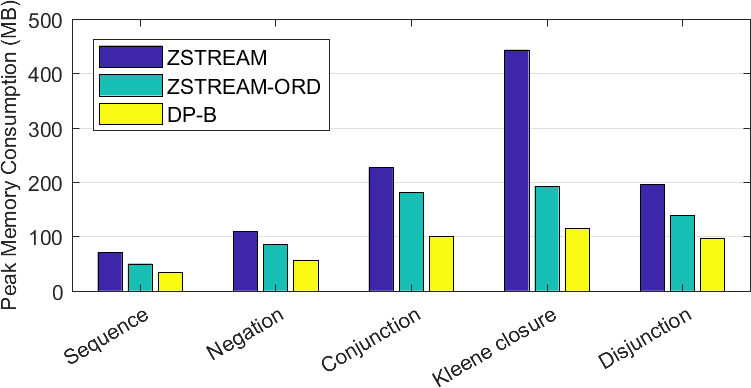}\label{fig:memory-tree}}
    \caption{Memory consumption for different pattern types (lower is better): \protect\subref{fig:memory-order} order-based methods; \protect\subref{fig:memory-tree} tree-based methods.}
	\label{fig:memory-by-type}
\end{figure}

Figures \ref{fig:throughput-by-type} and \ref{fig:memory-by-type}
present the comparison of the plan generation algorithms described
in Section \ref{sub:Evaluation-Algorithms} in terms of throughput
and memory consumption, respectively. Each group represents the results
obtained on a particular set of patterns described above, and each
bar depicts the average value of a performance metric for a particular
algorithm. For clarity, order-based and tree-based methods are shown
separately.

On average, the plans generated using JQPG algorithms achieve a considerably
higher throughput than those created using native CPG methods. For
order-based plans, the perceived gain of the best-performed DP-LD
over EFREQ ranged from a factor of 1.7 for iteration patterns to 2.7
for conjunctions. Similar results were obtained for tree-based plans
(ZSTREAM vs. DP-B). JQPG methods also display better overall memory
utilization. The order-based JQPG plans consume about 65-85\% of the
memory required by those produced by EFREQ. An even greater difference
was observed for tree-based plans, with DP-B using up to almost 4
times less memory than the CEP-native ZSTREAM.

Unsurprisingly, the best performance was observed for plans created
using the exhaustive algorithms based on dynamic programming, namely
DP-LD and DP-B. However, due to the exponential complexity of these
algorithms, their use in practice may be problematic for large patterns,
especially in systems where new evaluation plans are to be generated
with high frequency. Thus, one goal of the experimental study was
to test the exhaustive JQPG methods against the nonexhaustive ones
(such as GREEDY and II algorithms) to see whether the performance
gain of the former category is worth the high plan generation cost.

For the order-based case, the answer is indeed negative, as the results
for DP-LD and the heuristic JQPG algorithms are comparable and no
significant advantage is achieved by the former. Due to the relatively
small size of the left-deep tree space, the heuristics usually succeed
in locating the globally optimal plan. Moreover, the II-GREEDY algorithm
generally produces plans that are slightly more memory-efficient.
This can be attributed to our cost model, which only counts the partial
matches, but does not capture the other factors such as the size of
the buffered events. The picture looks entirely different for the
tree-based methods, where DP-B displays a convincing advantage over
both the basic ZStream algorithm and its combination with the greedy
heuristic method.

Another important conclusion from Figures \ref{fig:throughput-by-type}
and \ref{fig:memory-by-type} is that methods following the tree-based
model greatly outperform the order-based ones, both in throughput
and memory consumption. This is not a surprising outcome, as the tree-based
algorithms are capable of creating a significantly larger space of
plans. However, the best order-based JQPG algorithm (DP-LD)
is comparable or even superior to the CPG-native ZStream in most settings.

\begin{figure}
	\centering
	\subfloat[]{\includegraphics[width=.6\linewidth]{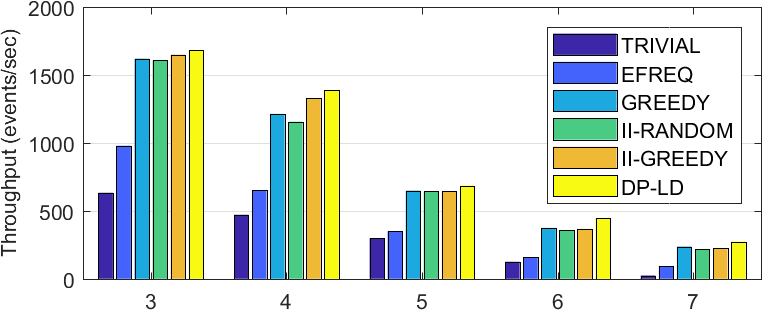}\label{fig:throughput-order}}\quad
	\subfloat[]{\includegraphics[width=.6\linewidth]{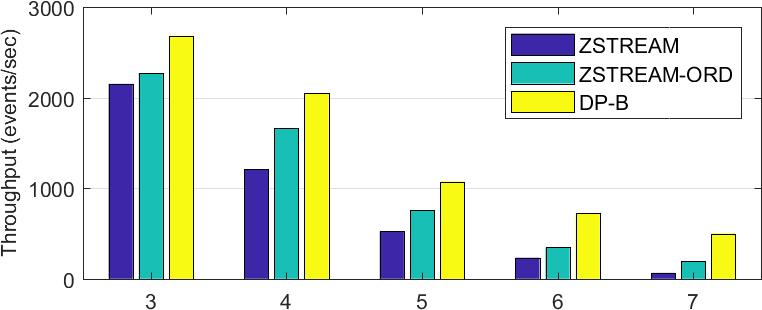}\label{fig:throughput-tree}}
    \caption{Throughput as a function of the sequence pattern size (higher is better): \protect\subref{fig:throughput-order} order-based methods; \protect\subref{fig:throughput-tree} tree-based methods.}
	\label{fig:sequence-throughput-by-size}
\end{figure}

\begin{figure}
	\centering
	\subfloat[]{\includegraphics[width=.6\linewidth]{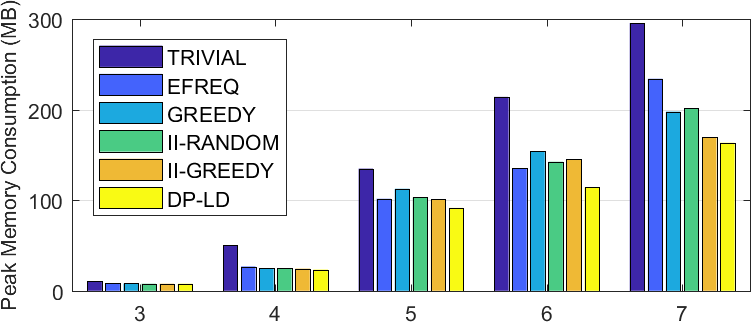}\label{fig:memory-order}}\quad
	\subfloat[]{\includegraphics[width=.6\linewidth]{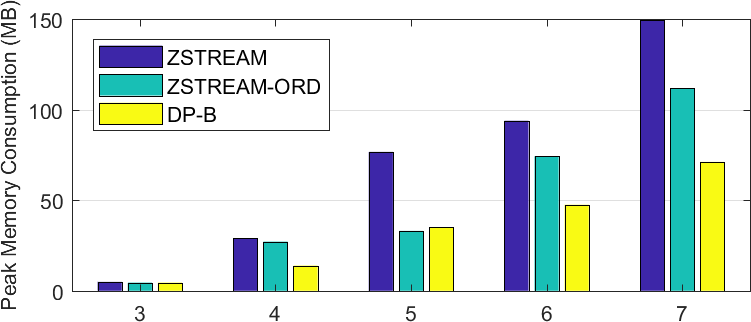}\label{fig:memory-tree}}
    \caption{Memory consumption as a function of the sequence pattern size (lower is better): \protect\subref{fig:memory-order} order-based methods; \protect\subref{fig:memory-tree} tree-based methods.}
	\label{fig:sequence-memory-by-size}
\end{figure}

\begin{figure}
	\centering
	\subfloat[]{\includegraphics[width=.6\linewidth]{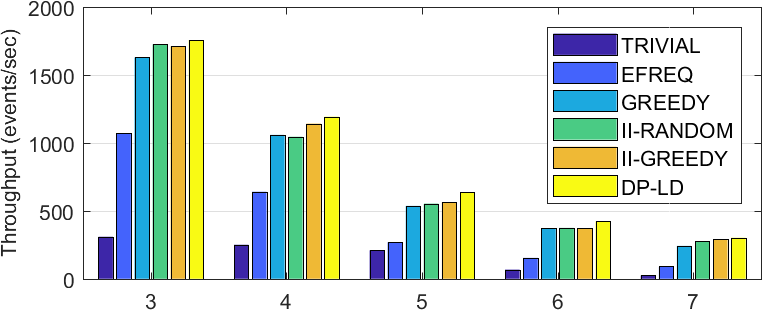}\label{fig:throughput-order}}\quad
	\subfloat[]{\includegraphics[width=.6\linewidth]{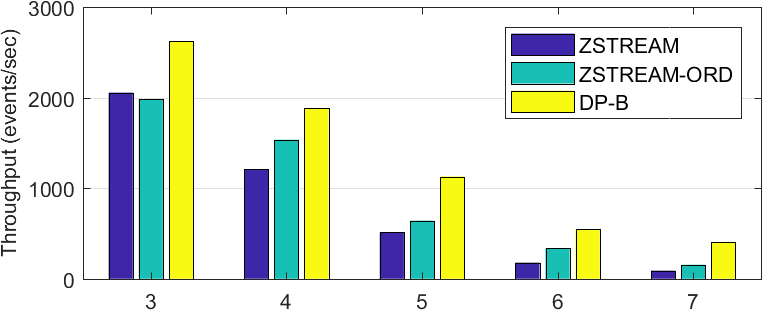}\label{fig:throughput-tree}}
    \caption{Throughput as a function of the negation pattern size (higher is better): \protect\subref{fig:throughput-order} order-based methods; \protect\subref{fig:throughput-tree} tree-based methods.}
	\label{fig:negation-throughput-by-size}
\end{figure}

\begin{figure}
	\centering
	\subfloat[]{\includegraphics[width=.6\linewidth]{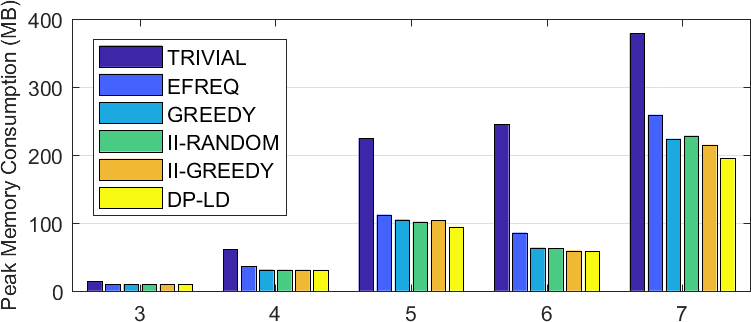}\label{fig:memory-order}}\quad
	\subfloat[]{\includegraphics[width=.6\linewidth]{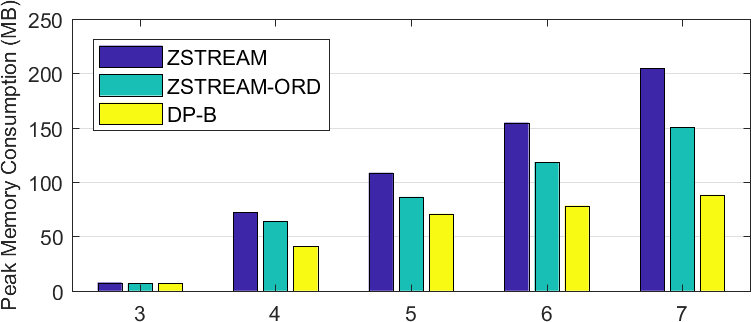}\label{fig:memory-tree}}
    \caption{Memory consumption as a function of the negation pattern size (lower is better): \protect\subref{fig:memory-order} order-based methods; \protect\subref{fig:memory-tree} tree-based methods.}
	\label{fig:negation-memory-by-size}
\end{figure}

\begin{figure}
	\centering
	\subfloat[]{\includegraphics[width=.6\linewidth]{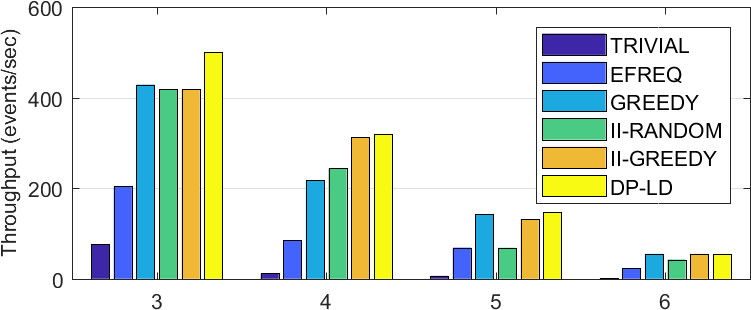}\label{fig:throughput-order}}\quad
	\subfloat[]{\includegraphics[width=.6\linewidth]{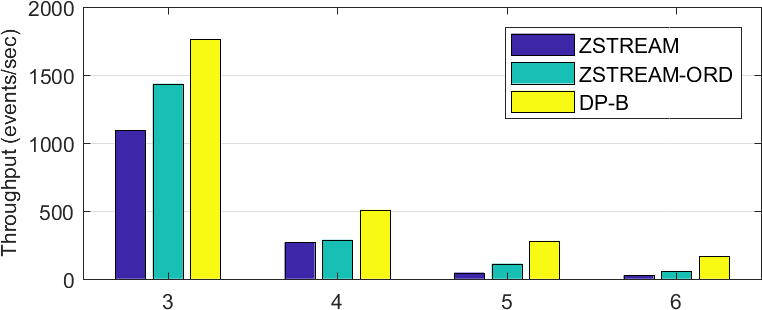}\label{fig:throughput-tree}}
    \caption{Throughput as a function of the conjunction pattern size (higher is better): \protect\subref{fig:throughput-order} order-based methods; \protect\subref{fig:throughput-tree} tree-based methods.}
	\label{fig:conjunction-throughput-by-size}
\end{figure}

\begin{figure}
	\centering
	\subfloat[]{\includegraphics[width=.6\linewidth]{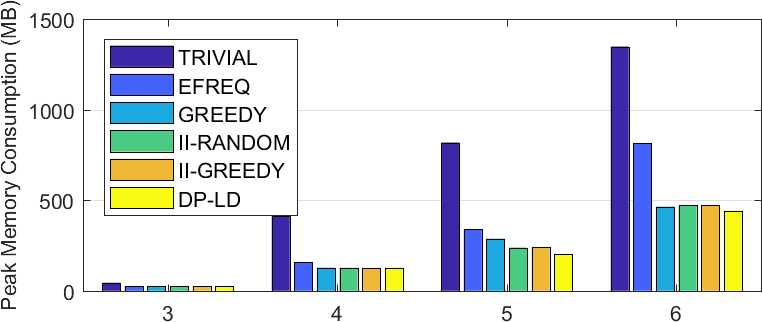}\label{fig:memory-order}}\quad
	\subfloat[]{\includegraphics[width=.6\linewidth]{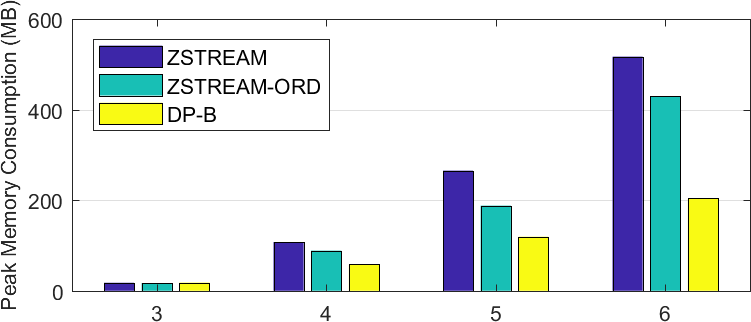}\label{fig:memory-tree}}
    \caption{Memory consumption as a function of the conjunction pattern size (lower is better): \protect\subref{fig:memory-order} order-based methods; \protect\subref{fig:memory-tree} tree-based methods.}
	\label{fig:conjunction-memory-by-size}
\end{figure}

\begin{figure}
	\centering
	\subfloat[]{\includegraphics[width=.6\linewidth]{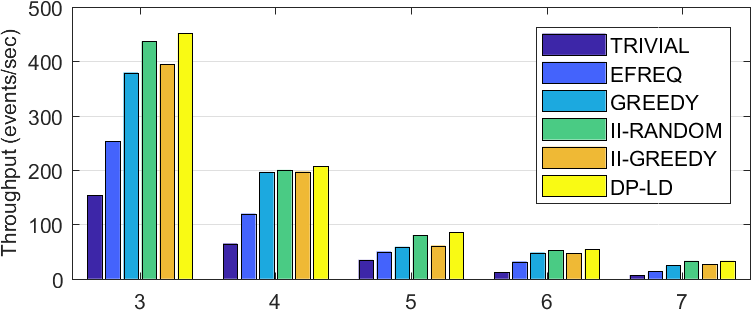}\label{fig:throughput-order}}\quad
	\subfloat[]{\includegraphics[width=.6\linewidth]{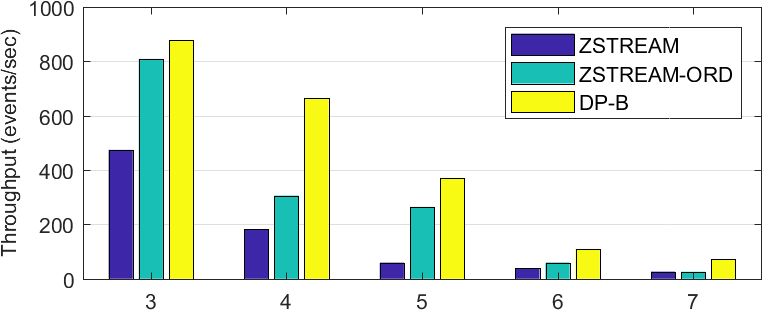}\label{fig:throughput-tree}}
    \caption{Throughput as a function of the iteration pattern size (higher is better): \protect\subref{fig:throughput-order} order-based methods; \protect\subref{fig:throughput-tree} tree-based methods.}
	\label{fig:iteration-throughput-by-size}
\end{figure}

\begin{figure}
	\centering
	\subfloat[]{\includegraphics[width=.6\linewidth]{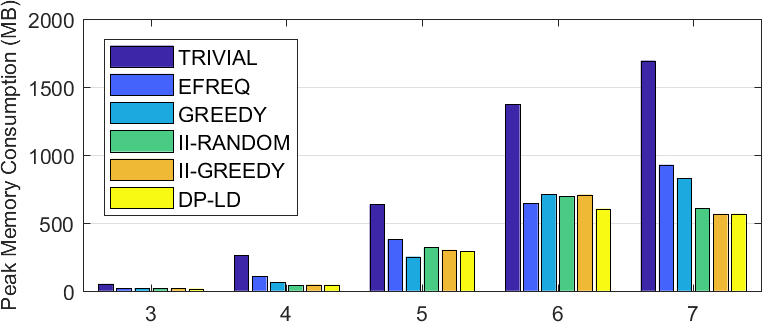}\label{fig:memory-order}}\quad
	\subfloat[]{\includegraphics[width=.6\linewidth]{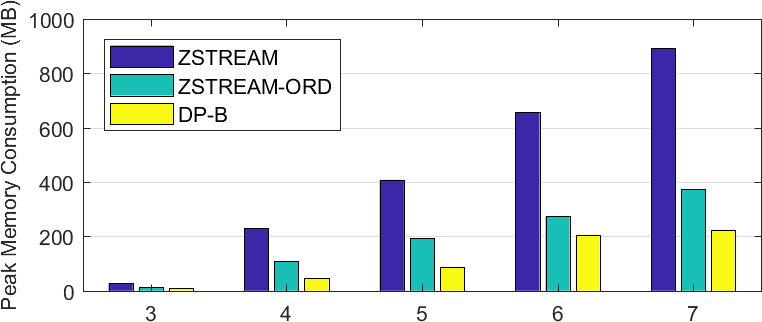}\label{fig:memory-tree}}
    \caption{Memory consumption as a function of the iteration pattern size (lower is better): \protect\subref{fig:memory-order} order-based methods; \protect\subref{fig:memory-tree} tree-based methods.}
	\label{fig:iteration-memory-by-size}
\end{figure}

\begin{figure}
	\centering
	\subfloat[]{\includegraphics[width=.6\linewidth]{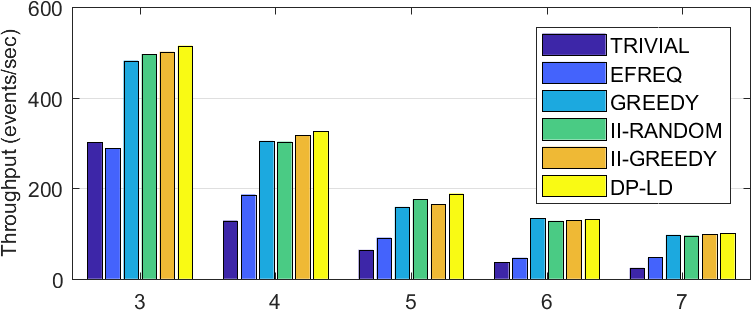}\label{fig:throughput-order}}\quad
	\subfloat[]{\includegraphics[width=.6\linewidth]{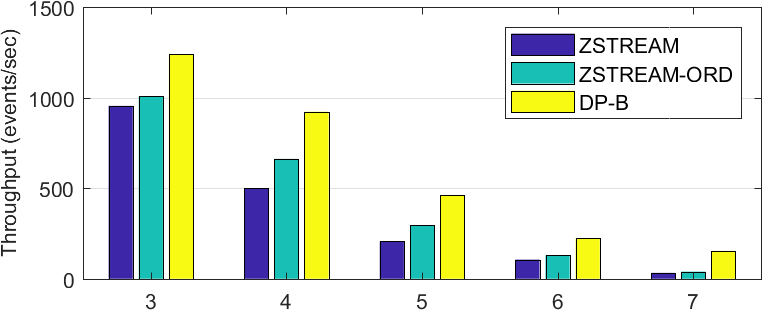}\label{fig:throughput-tree}}
    \caption{Throughput as a function of the disjunction pattern size (higher is better): \protect\subref{fig:throughput-order} order-based methods; \protect\subref{fig:throughput-tree} tree-based methods.}
	\label{fig:disjunction-throughput-by-size}
\end{figure}

\begin{figure}
	\centering
	\subfloat[]{\includegraphics[width=.6\linewidth]{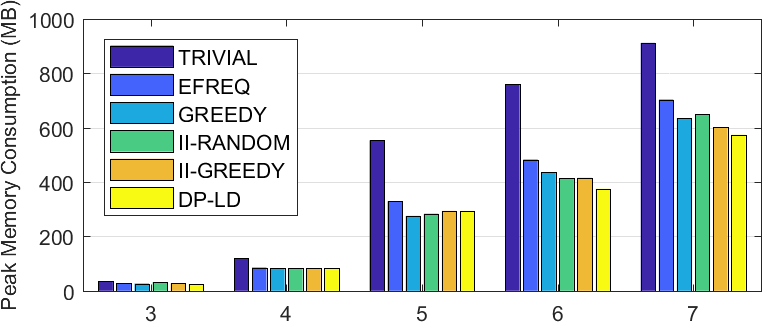}\label{fig:memory-order}}\quad
	\subfloat[]{\includegraphics[width=.6\linewidth]{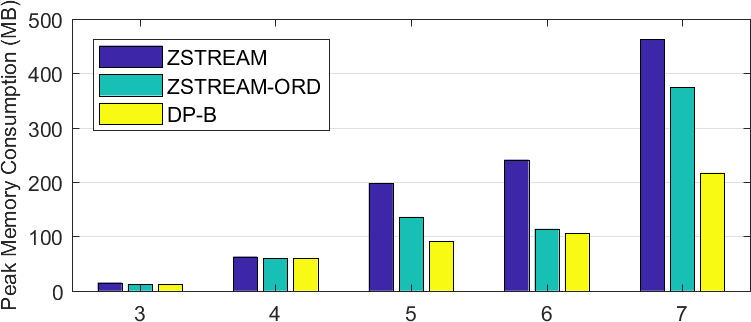}\label{fig:memory-tree}}
    \caption{Memory consumption as a function of the disjunction pattern size (lower is better): \protect\subref{fig:memory-order} order-based methods; \protect\subref{fig:memory-tree} tree-based methods.}
	\label{fig:disjunction-memory-by-size}
\end{figure}

Figures \ref{fig:sequence-throughput-by-size}-\ref{fig:disjunction-memory-by-size}
depict the results discussed above, partitioned by the pattern size.
Throughput for each of the five pattern categories described above
is displayed in Figures \ref{fig:sequence-throughput-by-size}, \ref{fig:negation-throughput-by-size},
\ref{fig:conjunction-throughput-by-size}, \ref{fig:iteration-throughput-by-size},
and \ref{fig:disjunction-throughput-by-size} respectively. Although
the performance of all methods degrades drastically as the pattern
size grows, the relative throughput gain for JQPG methods over native
CPG methods is consistently higher for longer sequences. This is especially
evident for the tree-based variation of the problem (\ref{fig:sequence-throughput-by-size}\subref{fig:throughput-tree}),
where the most efficient JQPG algorithm (DP-B) achieves 7.6 times
higher throughput than the native CPG framework (ZSTREAM) for patterns
of length 7, compared to a speedup of only 1.2 times for patterns
of 3 events. The results for memory consumption follow the same trend
(Figures \ref{fig:sequence-memory-by-size}, \ref{fig:negation-memory-by-size},
\ref{fig:conjunction-memory-by-size}, \ref{fig:iteration-memory-by-size},
and \ref{fig:disjunction-memory-by-size}). We can thus conclude that,
at least for the pattern sizes considered in this study, the JQPG
methods provide a considerably more scalable solution.

\begin{figure}
	\centering
	\subfloat[]{\includegraphics[width=\linewidth]{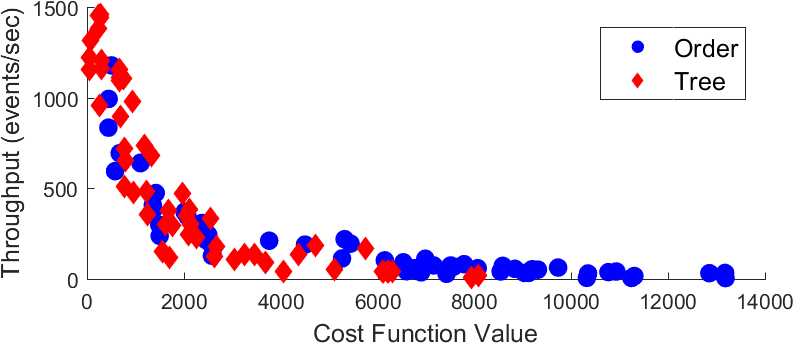}\label{fig:throughput-cost}}\quad
	\subfloat[]{\includegraphics[width=\linewidth]{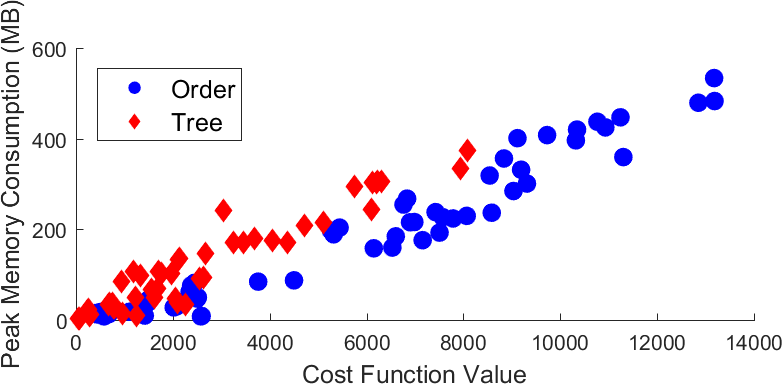}\label{fig:memory-cost}}
    \caption{Performance metrics as a function of the cost computed by the cost functions for order-based and tree-based patterns: \protect\subref{fig:throughput-cost} throughput; \protect\subref{fig:memory-cost} memory consumption.}
	\label{fig:cost-functions}
\end{figure}

In our next experiment, we evaluated the quality of the cost functions
used during plan generation. To that end, we created 60 order-based
and 60 tree-based plans for patterns of various types using different
algorithms. The plans were then executed on the stock dataset. The
throughput and the memory consumption measured during each execution
are shown in Figure \ref{fig:cost-functions} as the function of the
cost assigned to each plan by the corresponding function ($Cost_{ord}$
or $Cost_{tree}$). The obtained throughput seems to be inversely
proportional to the cost, behaving roughly as $\frac{1}{x^{c}};c\geq1$.
For memory consumption, an approximately linear dependency can be
observed. These results match our expectations, as a cheaper plan
is supposed to yield better performance and require less memory.
We may thus conclude that the costs returned by $Cost_{ord}$
and $Cost_{tree}$ provide a reasonably accurate estimation of the
actual performance of a plan.

The above conclusion allowed us to repeat the experiments summarized
in Figures \ref{fig:sequence-throughput-by-size}-\ref{fig:disjunction-memory-by-size}
for larger patterns, using the plan cost as the objective function. We generated 200 patterns
of sizes ranging from 3 to 22. We then created a set of plans for
each pattern using different algorithms and recorded the resulting
plan costs. Due to the exponential growth of the cost with the pattern
size, directly comparing the costs was impractical. Instead, the \textit{normalized
cost} was calculated for every plan. The normalized cost of a plan
$Pl$ created by an algorithm $A$ for a pattern $P$ was defined
as the cost of a plan generated for $P$ by the empirically worst
algorithm (the CEP-native EFREQ), divided by the cost of $Pl$.

\begin{figure}
	\centering
	\subfloat[]{\includegraphics[width=\linewidth]{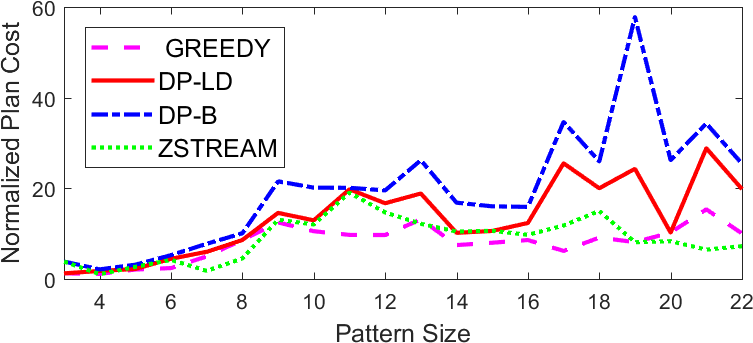}\label{fig:cost-value}}\quad
	\subfloat[]{\includegraphics[width=\linewidth]{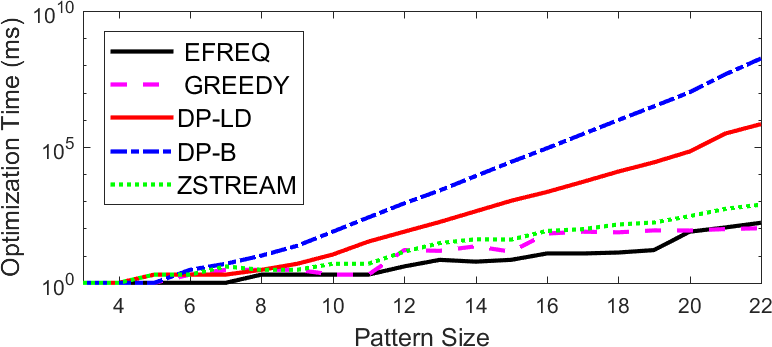}\label{fig:optimization-time}}
    \caption{Generation of large plans (selected algorithms): \protect\subref{fig:cost-value} average normalized plan cost (higher is better); \protect\subref{fig:optimization-time} average plan generation time (logarithmic scale, lower is better). The results are presented  as a function of pattern size.}
	\label{fig:large-patterns}
\end{figure}

The results for selected algorithms are depicted in Figure \ref{fig:large-patterns}\subref{fig:cost-value}.
Each data point represents an average normalized cost for all plans
of the same size created by the same algorithm. As we observed previously,
the DP-based join algorithms consistently produced significantly
cheaper plans (up to a factor of 57) than the heuristic alternatives.
Also, the worst JQPG method (GREEDY) and the best CPG method (ZSTREAM)
produced plans of similar quality, with the former slightly overperforming
the latter for larger pattern sizes. The worst-performing EFREQ algorithm
was used for normalized cost calculation and is thus not shown in
the figure.

Figure \ref{fig:large-patterns}\subref{fig:optimization-time} presents
the plan generation times measured during the above experiment. The
results are displayed in logarithmic scale. While all algorithms
incur only negligible optimization overhead for small patterns, it grows
rapidly for methods based on dynamic programming (for a pattern of
length 22, it took over 50 hours to create a plan using DP-B). This
severely limits the applicability of the DP-based approaches when
the number of events in a pattern is high. On the other hand, all
non-dynamic algorithms were able to complete in under a second even
for the largest tested patterns. The join-based greedy algorithm (GREEDY)
demonstrated the best overall trade-off between optimization time
and quality.

Next, we studied the performance of the hybrid throughput-latency
cost model introduced in Section \ref{sub:Pattern-Detection-Latency}.
Each of the 6 JQPG-based methods discussed in Section \ref{sub:Evaluation-Algorithms}
was evaluated using three different values for the throughput-latency
trade-off parameter $\alpha$: 0, 0.5 and 1. Note that for the first
case ($\alpha=0$) the resulting cost model is identical to the one
defined in Section \ref{sec:The-Equivalence-of} and used in the experiments
above. For each algorithm and for each value of $\alpha$, the throughput
and the average latency (in milliseconds) were measured.

\begin{figure}
	\centering
	\includegraphics[width=.7\linewidth]{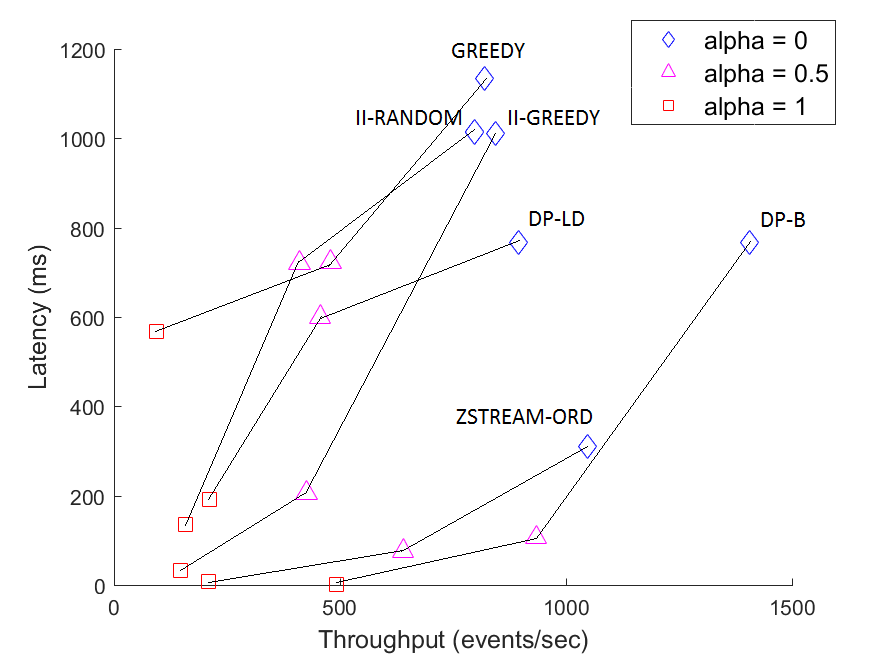}
    \caption{Throughput vs. latency using different values for the alpha parameter of the cost model.}
	\label{fig:latency}
\end{figure}

Figure \ref{fig:latency} demonstrates the results, averaged over
500 patterns included in the sequence pattern set. Measurements obtained
using the same algorithm are connected by straight lines, and the
labels near the highest points (diamonds) indicate the algorithms
corresponding to these points. It can be seen that increasing the
value of $\alpha$ results in a significantly lower latency. However,
this also results in a considerable drop in throughput for most algorithms.
By fine-tuning this parameter, the desired latency can be achieved
with minimal loss in throughput. It can also be observed that the
tree-based algorithms DP-B and ZSTREAM-ORD (and, to some extent, the
order-based II-GREEDY) achieve a substantially better throughput-latency
trade-off as compared to other methods.

\begin{figure}
	\centering
	\subfloat[]{\includegraphics[width=.6\linewidth]{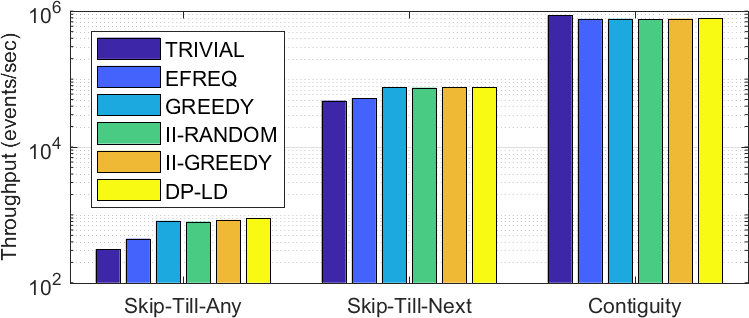}\label{fig:strategy-order}}\quad
	\subfloat[]{\includegraphics[width=.6\linewidth]{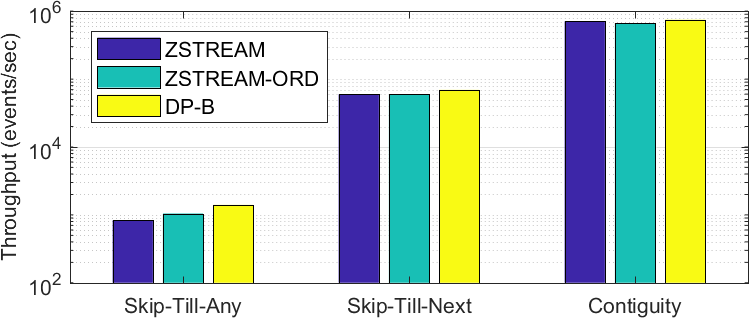}\label{fig:strategy-tree}}
    \caption{Throughput for different event selection strategies (logarithmic scale): \protect\subref{fig:strategy-order} order-based methods; \protect\subref{fig:strategy-tree} tree-based methods.}
	\label{fig:throughput-by-selection-strategy}
\end{figure}

Finally, we performed a comparative throughput evaluation of the sequence
pattern set under three different event selection strategies: \textit{skip-till-any-match},
\textit{skip-till-next-match} and \textit{contiguity} (Section \ref{sub:Event-Consumption-Policies}).
The results are depicted in Figure \ref{fig:throughput-by-selection-strategy}
for all algorithms under examination. Due to large performance gaps
between the examined methods, the results are displayed in logarithmic
scale.

For \textit{skip-till-next-match}, JQPG methods hold a clear advantage,
albeit less significant than the one demonstrated above for \textit{skip-till-any-match}.
The opposite observation can be made about the \textit{contiguity}
strategy, where the trivial algorithm following a static plan outperforms
other, more complicated methods. Due to the simplicity of the event
detection process and the lack of nondeterminism in this case, the
plan set by an input specification always performs best, while the
alternatives introduce a slight additional overhead of reordering
and event buffering.

\section{Related Work}

\label{sec:Related-Work}

Systems for scalable extraction of complex events from high-speed
information flows have become an increasingly important research field
during last decades, as a result of the rising demand for technologies
of this type \cite{CugolaM12,EtzionN10}. Their inception can be traced
to earlier systems for massive data stream processing, such as TelegraphCQ \cite{ChandrasekaranDFHHKMRRS03},
NiagaraCQ \cite{ChenDTW00}, Aurora/Borealis \cite{AbadiABCHLMRRTXZ05}
and Stream \cite{ArasuBBCDIMSW16}. Later, a broad variety of general
purpose complex event processing solutions emerged \cite{AdiE04,AkdereMCT08,BargaGAH07,CugolaGM12,DemersJB07,KolchinskySS15,LiuRDGWAM11,MeiM09,RabinovichEG11,Schultz-MollerMP09,WuDR06},
including the widely used commercial CEP providers, such as Esper \cite{Esper10}
and IBM System S \cite{AminiABEKSPV06}.

Various performance optimization techniques are implemented in complex
event processing systems \cite{HirzelSSGG14}. In \cite{RabinovichEG11}, a rewriting framework
is described, based on unifying and splitting patterns. A method for
efficient Kleene closure evaluation based on sharing with postponed
operators is discussed in \cite{ZhangDI2014}, while in \cite{PoppeLAR17}
the authors solve the above problem by maintaining a compact graph
encoding of event sequences and utilizing it for effective reuse.
RunSAT \cite{DingCRTHS08} utilizes another approach, preprocessing
a pattern and setting optimal points for termination of the detection
process. ZStream \cite{MeiM09} presents an optimization framework
for optimal tree generation, based on a complex cost model. As already
shown above, since the leaves of an evaluation tree cannot be reordered,
it only searches through a partial solution space. Advanced methods were also 
proposed for multi-query CEP optimization \cite{DemersGHRW06,LiuRGGWAM11,RayLR16,RayRLGWA13,ZhangVDH17}.

CEP engines utilizing the order-based evaluation approach have also
adopted different optimization strategies. SASE \cite{WuDR06}, Cayuga \cite{DemersJB07}
and T-Rex \cite{CugolaGM12} design efficient data structures to enable
smart runtime memory management. These NFA-based mechanisms do not
support out-of-order processing, and hence are still vulnerable to
the problem of large intermediate results. In \cite{AkdereMCT08,KolchinskySS15,Schultz-MollerMP09},
various pattern reordering methods for efficient order-based complex
event detection are described. None of these works takes the selectivities
of the event constraints into account.

Estimating an optimal evaluation plan for a join query has long been
considered one of the most important problems in the area of query
optimization \cite{SteinbrunnMK97}. Multiple authors have shown the
NP-completeness of this problem for arbitrary query graphs \cite{CluetM95,Ibaraki84},
and a wide range of methods were proposed to provide either exact
or approximate close-to-optimal solutions \cite{KossmannS00,KrishnamurthyBZ86,LeeSC97,MoerkotteN06,SelingerACLP79,SteinbrunnMK97,Swami89}.

Methods for join query plan generation can be roughly divided into
two main categories. The heuristic algorithms, as their name suggests,
utilize some kind of heuristic function or approach to perform efficient
search through the huge solution space. They are often applied in
conjunction with combinatorial \cite{IoannidisK90,SteinbrunnMK97,Swami89}
or graph-based \cite{KrishnamurthyBZ86,LeeSC97} techniques. The heuristic
algorithms produce fast solutions, but the resulting execution plans
are often far from the optimum.

The second category of JQPG algorithms, the exhaustive search algorithms,
provide provable guarantees on the optimality of the returned solutions.
These methods are often based on dynamic programming \cite{MoerkotteN06,SelingerACLP79}
and thus suffer from worst-case exponential complexity. To solve this
issue, hybrid techniques were proposed, making it possible to set
the trade-off between the speed of heuristic approaches and the precision
of DP-based approaches \cite{KossmannS00}.

Incorporating join optimization techniques from traditional DBMSs
was already considered in the related fields, such as XPath \cite{GrustRT07}
and data stream processing \cite{ChenDTW00}. For the best of our
knowledge, our work is the first to address the CEP-specific challenges
and to provide a formal reduction.

\section{Conclusions and Future Work}

\label{sec:Conclusions-and-Future}

In this paper, we studied the relationship between two important and
relevant problems, CEP Plan Generation and Join Query Plan Generation.
It was shown that the CPG problem is equivalent to JQPG for a subset
of pattern types, and reducible to it for other types. We discussed
how close-to-optimal solutions to CPG can be efficiently obtained
by applying existing JQPG methods. CEP-related challenges, such as
detection latency and event selection strategies, were addressed.
The presented experimental study supported our analysis by demonstrating
how the evaluation plans created by some of the well-known join algorithms
outperform those produced by the methods traditionally used in CEP
systems.

Utilizing join-related techniques in the field of CEP introduces additional,
not yet addressed challenges, such as efficient tracking of real-time
predicate selectivities and handling inter-predicate dependencies.
We intend to target these challenges in our future research, in 
addition to the directions outlined throughout the paper.

\bibliographystyle{plain}
\bibliography{references}

\appendix

\section{ASI Property of the Order-Based Cost Functions}

\label{sec:ASI-Property-of}

In this appendix, we will formally prove that the order-based CPG
cost functions $Cost_{ord}^{trpt}$ and $Cost_{ord}^{lat}$ presented
in Sections \ref{sub:Order-Based-Evaluation} and \ref{sub:Pattern-Detection-Latency}
respectively have the adjacent sequence interchange (ASI) property,
defined in \cite{MonmaS79}. As we discussed in Section \ref{sub:Discussion},
polynomial-time algorithms were developed for join ordering of acyclic
queries subject to cost functions that have this property \cite{Ibaraki84,KrishnamurthyBZ86}.
Since all JQPG algorithms demonstrated in this paper are executed
subject to the left-deep tree cost function $Cost_{LDJ}$, we do not
use this result directly for solving the CPG problem. However, the
throughput- and latency-related functions can potentially be employed
directly for solving the join ordering problem in streaming database
systems, and hence it is important to show their ASI property as a
part of our work.

We will start with the definition of the ASI property.

\newtheorem{definition}{Definition}
\begin{definition}

A cost function $C$ has the adjacent sequence interchange (ASI) property,
if and only if there exists a rank function $rank\left(s\right)$
for sequences $s$, such that for all sequences $a,b$ and for all
non-empty sequences $v,u$ the following holds:
\[
C\left(auvb\right)\leq C\left(avub\right)\Leftrightarrow rank\left(u\right)\leq rank\left(v\right).
\]

\end{definition}

We will first provide the proof for the throughput-related cost function
$Cost_{ord}^{trpt}$ . To that end, we will utilize the idea from
a similar proof in \cite{CluetM95}.

\begin{thrm}

The cost function $Cost_{ord}^{trpt}$ has the ASI property.

\end{thrm}

Let $P$ be a pure conjunctive pattern over the event types $T_{1},\cdots,T_{n}$
with an acyclic query graph. Recall that the cost function is defined
as follows:
\[
Cost_{ord}^{trpt}\left(O\right)=\sum_{k=1}^{n}\left(W^{k}\cdot\prod_{i=1}^{k}r_{p_{i}}\cdot\prod_{i,j\leq k;i\leq j}sel_{p_{i},p_{j}}\right),
\]
where $O=\left(T_{p_{1}},T_{p_{2}},\cdots T_{p_{n}}\right);p_{i}\in[1,n]$.

Due to the acyclicity of the pattern, each event type $T_{p_{i}}$
will only have one predicate with the event types preceding it in
$O$. Further, if we set the root of the query tree at some event
type $T_{R}$, this predicate can be uniquely determined for any other
type, as follows from the uniqueness of a path between two nodes in
a tree. For each $T_{i}\neq T_{R}$, we will denote this predicate
as $c_{i}^{R}$, and its selectivity will be thus denoted as $sel_{i}^{R}$.
For the root event type, we set $sel_{R}^{R}=1$. Rewriting the cost
function definition accordingly, we get:
\[
Cost_{ord}^{trpt}\left(O\right)=\sum_{k=1}^{n}\prod_{i=1}^{k}\left(W\cdot r_{p_{i}}\cdot sel_{i}^{R}\right).
\]

We will now define the following auxiliary functions, defined on any
sequence $s$ of size $m$, consisting of events $T_{1},\cdots,T_{n}$:
\[
C\left(s\right)=\sum_{k=1}^{m}\prod_{i=1}^{k}\left(W\cdot r_{p_{i}}\cdot sel_{i}^{R}\right);\: C\left(\varepsilon\right)=0
\]
\[
T\left(s\right)=\prod_{i=1}^{m}\left(W\cdot r_{p_{i}}\cdot sel_{i}^{R}\right);\: T\left(\varepsilon\right)=1.
\]

Note that the above functions only depend on the selection of the
root $R$.

It can be observed that $C\left(O\right)=Cost_{ord}^{trpt}\left(O\right)$
for any evaluation order $O$. In addition, the following holds:
\[
C\left(s_{1}s_{2}\right)=C\left(s_{1}\right)+T\left(s_{1}\right)\cdot C\left(s_{2}\right).
\]

We will now define the rank function as follows:
\[
rank\left(s\right)=\frac{T\left(s\right)-1}{C\left(s\right)}
\]

We will demonstrate the ASI property for the rank function $rank$.
Let $a,b$ arbitrary sequences and $v,u$ arbitrary non-empty sequences.
Then:

\begin{align*}
C\left(auvb\right)&\leq C\left(avub\right)\Leftrightarrow\\
C\left(a\right)+T\left(a\right)\cdot C\left(uvb\right)&\leq C\left(a\right)+T\left(a\right)\cdot C\left(vub\right)\Leftrightarrow\\
C\left(u\right)+T\left(u\right)\cdot C\left(vb\right)&\leq C\left(v\right)+T\left(v\right)\cdot C\left(ub\right)\Leftrightarrow\\
C\left(u\right)+T\left(u\right)\cdot\left(C\left(v\right)+T\left(v\right)\cdot C\left(b\right)\right)&\leq C\left(v\right)+T\left(v\right)\cdot\left(C\left(u\right)+T\left(u\right)\cdot C\left(vb\right)\right)\Leftrightarrow\\
C\left(u\right)+T\left(u\right)\cdot C\left(v\right)+T\left(u\right)\cdot T\left(v\right)\cdot C\left(b\right)&\leq C\left(v\right)+T\left(v\right)\cdot C\left(u\right)+T\left(v\right)\cdot T\left(u\right)\cdot C\left(b\right)\Leftrightarrow \\
T\left(u\right)\cdot C\left(v\right)-C\left(v\right)&\leq T\left(v\right)\cdot C\left(u\right)-C\left(u\right)\Leftrightarrow\\
\frac{T\left(u\right)-1}{C\left(u\right)}&\leq\frac{T\left(v\right)-1}{C\left(v\right)}\Leftrightarrow\\
rank\left(u\right)&\leq rank\left(v\right).
\end{align*}

The above transitions are possible due to the positivity of $T\left(s\right)$
and the positivity of $C\left(s\right)$ for non-empty sequences.$\blacksquare$

We will now proceed to the latency-related cost function $Cost_{ord}^{lat}$
.

\begin{thrm}

The cost function $Cost_{ord}^{lat}$ has the ASI property.

\end{thrm}

Let $P$ be a pure conjunctive pattern over the event types $T_{1},\cdots,T_{n}$
with an acyclic query graph. Recall that the cost function is defined
as follows:
\[
Cost_{ord}^{lat}\left(O\right)=\sum_{T_{i}\in Succ_{O}\left(T_{n}\right)}W\cdot r_{i},
\]
where $O=\left(T_{p_{1}},T_{p_{2}},\cdots T_{p_{n}}\right);p_{i}\in[1,n]$,
$T_{n}$ is the last event type in the order induced by the pattern,
and $Succ_{O}\left(T_{n}\right)$ denotes the event types succeeding
$T_{n}$ in $O$.

For each sequence $s$ over the above types, we will define the following
function accepting an event type $T_{i}$ as a parameter:
\[
g_{s}\left(T_{i}\right)=\begin{cases}
W\cdot r_{i} & if\: T_{i}\in Succ_{s}\left(T_{n}\right)\\
0 & otherwise.
\end{cases}
\]

$Cost_{ord}^{lat}$ can be now rewritten in the following form:
\[
Cost_{ord}^{lat}\left(O\right)=\sum_{i=1}^{n}g_{O}\left(T_{i}\right).
\]

The rank function will be defined as follows:
\[
rank\left(s\right)=\begin{cases}
\sum_{T_{i}\in s}g_{s}\left(T_{i}\right) & if\: T_{n}\in s\\
0 & otherwise.
\end{cases}
\]

We will now demonstrate the ASI property of $Cost_{ord}^{lat}$ by
examining the following cases for non-empty sequences $u,v$:
\begin{enumerate}
\item $T_{n}\notin u$ and $T_{n}\notin v$: in this case, $rank\left(u\right)=rank\left(v\right)=0$.
Examine the expressions $Cost_{ord}^{lat}\left(auvb\right)$ and $Cost_{ord}^{lat}\left(avub\right)$
for some sequences $a,b$. If $T_{n}\notin a$, then obviously $u$
and $v$ do not contribute any event type with a non-zero value of
$g_{auvb}\left(T_{i}\right)$ (or $g_{avub}\left(T_{i}\right)$, symmetrically),
hence $Cost_{ord}^{lat}\left(auvb\right)=Cost_{ord}^{lat}\left(avub\right)=Cost_{ord}^{lat}\left(ab\right)$.
Otherwise, since $u,v,b\subseteq Succ_{auvb}\left(T_{n}\right)$,
the cost function for $u,v,b$ is commutative by definition and thus
$Cost_{ord}^{lat}\left(auvb\right)=Cost_{ord}^{lat}\left(avub\right)$.
\item $T_{n}\in v$: in this case, we have $rank\left(v\right)\geq0$ by
non-negativity of $g_{s}$ and $rank\left(u\right)=0$ since $u$
and $v$ must be disjoint. Consequently, $rank\left(u\right)\leq rank\left(v\right)$.
Now, examine the first cost expression:\begin{align*}
&Cost_{ord}^{lat}\left(auvb\right)=\sum_{T_{i}\in auvb}g_{auvb}\left(T_{i}\right)=\\
&=\sum_{T_{i}\in a}g_{auvb}\left(T_{i}\right)+\sum_{T_{i}\in u}g_{auvb}\left(T_{i}\right)+\sum_{T_{i}\in v}g_{auvb}\left(T_{i}\right)+\sum_{T_{i}\in b}g_{auvb}\left(T_{i}\right)=\\
&=0+0+rank\left(v\right)+\sum_{T_{i}\in b}\left(W\cdot r_{i}\right).
\end{align*}%
For the second cost expression, we obtain the following:\begin{align*}
&Cost_{ord}^{lat}\left(avub\right)=\sum_{T_{i}\in avub}g_{avub}\left(T_{i}\right)=\\
&=\sum_{T_{i}\in a}g_{avub}\left(T_{i}\right)+\sum_{T_{i}\in v}g_{avub}\left(T_{i}\right)+\sum_{T_{i}\in u}g_{avub}\left(T_{i}\right)+\sum_{T_{i}\in b}g_{avub}\left(T_{i}\right)=\\
&=0+rank\left(v\right)+\sum_{T_{i}\in u}\left(W\cdot r_{i}\right)+\sum_{T_{i}\in b}\left(W\cdot r_{i}\right).
\end{align*}%
The difference between these expressions is:
\[
Cost_{ord}^{lat}\left(avub\right)-Cost_{ord}^{lat}\left(auvb\right)=\sum_{T_{i}\in u}\left(W\cdot r_{i}\right).
\]
By non-negativity of time windows and arrival rates, we thus get $Cost_{ord}^{lat}\left(auvb\right)\leq Cost_{ord}^{lat}\left(avub\right)$.
\item $T_{n}\in u$: this case is symmetrical to case 2, and identical steps
are applied to show that $rank\left(v\right)\leq rank\left(u\right)$
and $Cost_{ord}^{lat}\left(avub\right)\leq Cost_{ord}^{lat}\left(auvb\right)$.
\end{enumerate}
To summarize, we have demonstrated that in all cases the condition
of the ASI property holds, which completes the proof.$\blacksquare$

\end{document}